\documentclass[sigconf]{acmart}

\usepackage{booktabs}
\usepackage{enumitem}
\usepackage{physics}
\usepackage{tabularx}
\usepackage{amsmath}
\usepackage{subcaption}
\usepackage{siunitx}
\usepackage{gensymb}
\usepackage{graphicx}
\usepackage{siunitx}

\copyrightyear{2025}
\acmYear{2025}
\setcopyright{cc}
\setcctype{by-sa}
\acmConference[SIGIR-AP 2025]{Proceedings of the 2025 Annual International ACM SIGIR Conference on Research and Development in Information Retrieval in the Asia Pacific Region}{December 7--10, 2025}{Xi'an, China}
\acmBooktitle{Proceedings of the 2025 Annual International ACM SIGIR Conference on Research and Development in Information Retrieval in the Asia Pacific Region (SIGIR-AP 2025), December 7--10, 2025, Xi'an, China}\acmDOI{10.1145/3767695.3769476}
\acmISBN{979-8-4007-2218-9/2025/12}

\begin{document}
\title{REGENT: Relevance-Guided Attention for Entity-Aware Multi-Vector Neural Re-Ranking}

\author{Shubham Chatterjee}
\orcid{0000-0002-6729-1346}
\affiliation{% 
    \institution{Missouri University of Science and Technology} 
    \department{Department of Computer Science} 
    \city{Rolla} 
    \state{Missouri} 
    \country{United States}
}
\email{shubham.chatterjee@mst.edu}




\begin{abstract}

% Current neural re-rankers perform well on short queries and passages but struggle with real-world retrieval tasks involving complex queries and long, information-rich documents. Unlike humans who rely on key entities to anchor understanding, neural models operate within fixed token windows and lack semantic guidance, often missing critical context. While attention mechanisms are designed to focus on important signals, existing models treat all token interactions as equally plausible. To bridge this gap, we introduce \textsc{REGENT}, a multi-vector entity-oriented neural re-ranking model built around a \emph{relevance-guided attention mechanism} that explicitly biases attention using both fine-grained lexical signals (via token-level BM25 scores) and query-specific entity semantics.

% Experiments on TREC Robust04, TREC Core18, and CODEC show that \textsc{REGENT} achieves state-of-the-art performance---delivering up to 108\% improvement over BM25 and outperforming strong neural and LLM-based baselines including RankT5, ColBERT, and RankVicuna. Ablation studies reveal a 74\% drop in performance when entities are removed, confirming their importance as a ``semantic skeleton'' for effective retrieval. To our knowledge, this is the first work to successfully integrate entity information to guide the attention mechanism, laying the groundwork for a new generation of entity-aware neural retrievers.

Current neural re-rankers often struggle with complex information needs and long, content-rich documents. The fundamental issue is not computational--it is \emph{intelligent content selection}: identifying what matters in lengthy, multi-faceted texts. While humans naturally anchor their understanding around key entities and concepts, neural models process text within rigid token windows, treating all interactions as equally important and missing critical semantic signals. We introduce \textsc{REGENT}, a neural re-ranking model that mimics human-like understanding by using entities as a ``semantic skeleton'' to guide attention. REGENT integrates relevance guidance directly into the attention mechanism, combining fine-grained lexical matching with high-level semantic reasoning. This relevance-guided attention enables the model to focus on conceptually important content while maintaining sensitivity to precise term matches. \textsc{REGENT} achieves new state-of-the-art performance in three challenging datasets, providing up to 108\% improvement over BM25 and consistently outperforming strong baselines including ColBERT and RankVicuna.  To our knowledge, this is the first work to successfully integrate entity semantics directly into neural attention, establishing a new paradigm for entity-aware information retrieval.

    %Neural re-ranking models have shown significant improvements in document ranking through their ability to capture complex semantic relationships. Recent work has demonstrated the benefits of incorporating traditional ranking signals like BM25 and leveraging entity knowledge from external knowledge bases. However, current approaches use these signals only at the document level or through precomputed entity similarity matrices, missing opportunities for fine-grained interaction between lexical matches and entity information. We present \texttt{REGENT}, a multi-vector neural re-ranking model that creates query-specific document representations by guiding its attention mechanism using both token-level BM25 scores and entity-aware attention pathways. 
    
    %Unlike previous approaches that rely on document-level signals or static entity similarity matrices, \texttt{REGENT} incorporates relevance signals directly into its attention computation through a dynamic fusion mechanism that learns to balance token-level matching with entity-level semantic understanding. Each document token's contribution to the final relevance score is weighted by both its query-specific BM25 score and its relationships with relevant entities. Through a multi-vector architecture, our model can process token and entity information at a fine granularity while maintaining efficiency. Experiments on standard benchmarks show that \texttt{REGENT} achieves state-of-the-art performance, particularly on difficult queries where traditional approaches fail, setting a foundation for future work in entity-aware retrieval.

\end{abstract}

\maketitle

\section{Introduction}
\label{sec:Introduction}

Neural retrieval models perform well on benchmarks with short, factual queries and brief passages~\cite{khattab2020colbert,karpukhin-etal-2020-dense,xiao-etal-2022-retromae}, achieving substantial gains over traditional methods on datasets such as MS MARCO~\cite{bajaj2016ms} and the TREC Deep Learning tracks~\cite{craswell2021trecdl}. However, real-world search often involves complex queries that require multi-hop reasoning, synthesis of evidence, or understanding relationships across long, information-rich documents such as news articles or scientific papers. For example, answering the query ``\textit{Was the crash that followed the dot-com bubble an overreaction considering the ultimate success of the Internet?}'' demands not just factual recall but also historical and interpretive reasoning. A human searching for an answer in a long article would naturally focus on themes like economic impact and the rise of internet businesses, while ignoring unrelated technical details. In contrast, neural models typically process only a limited portion of text (e.g., first 512 tokens in BERT) or divide documents into chunks that are scored independently before aggregation~\cite{li2020parade}. Each of these approaches risks missing globally relevant context, which can lead to degraded performance on complex queries.

We posit that the core challenge is not simply handling longer input, but performing \emph{intelligent content selection}: identifying and focusing on the most relevant portions of complex documents. We argue that entities can play a central role in addressing this gap.  Entities can be viewed as a high-level \textit{semantic skeleton} that exposes the conceptual structure of a document. Entities can segment text into regions related to key topics or actors, providing a coarse but meaningful map of the content. For example, in a discussion of the dot-com bubble, entities like ``\texttt{Nasdaq},'' ``\texttt{Amazon},'' and ``\texttt{venture capital}'' highlight market dynamics and investment trends. Crucially, entity representations are computed independently of token windows, enabling models to capture salient context beyond local spans. Existing entity-aware methods~\cite{xiong2017end,liu-etal-2018-entity} typically rely on static, context-agnostic entity similarity matrices, treating all entities as equally important regardless of query. This does not capture the dynamic semantic alignment between the query and document entities that is essential for modeling complex queries. For instance, ``\texttt{Amazon}'' might be highly relevant for queries about e-commerce success stories, but less important for queries focused on the broader economic policy implications of the dot-com era.

\begin{figure*}[t]
    \centering
    \includegraphics [width=\textwidth]{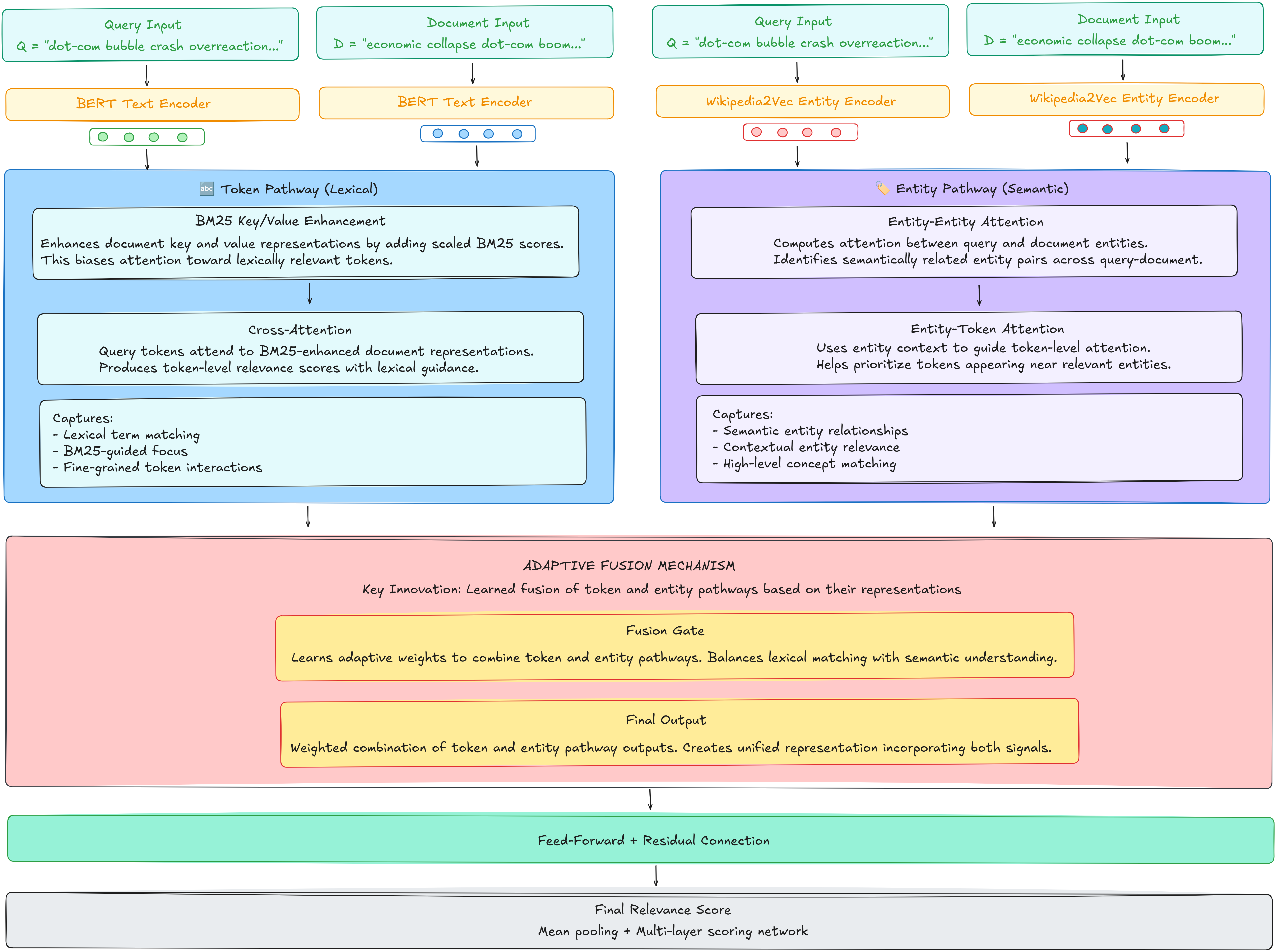}
    \caption{REGENT Architecture Overview. The model processes query and document inputs through separate BERT encoders to generate contextual embeddings. REGENT employs a dual-pathway attention mechanism: (1) The token pathway enhances document key and value representations with BM25 scores, then applies cross-attention to capture lexical matches between query and document tokens. (2) The entity pathway first projects pre-computed entity embeddings to the hidden dimension, computes entity-entity attention to identify semantically related concepts, then uses this entity context to guide token-level attention. An adaptive fusion mechanism learns to combine both pathways, balancing lexical matching with semantic understanding. The final output undergoes feed-forward processing with residual connections before mean pooling and scoring to produce a relevance score. This architecture enables fine-grained integration of traditional IR signals (BM25) with neural semantic reasoning through entities, moving beyond post-hoc score combination to embedded relevance guidance within the attention mechanism itself.
    }
    \label{fig:regent-architecture}
\end{figure*}

In parallel, lexical signals (e.g., BM25), remain a strong source of fine-grained evidence, offering cues based on exact term overlap. Previous hybrid approaches have attempted to combine lexical and neural signals~\cite{askari2023injecting,wang2021bert}, but typically do so post hoc, merging independently computed document level scores. This late stage fusion treats lexical and semantic signals as separate streams, preventing lexical cues from influencing the neural model's internal representation learning. We argue that lexical signals such as BM25 should not simply be combined with neural output, but should \emph{actively guide the neural ranking process itself}. This requires moving beyond document-level fusion and \emph{integrating relevance signals directly into the attention mechanism}, where neural models form their understanding of query-document relationships. Since attention operates at the token level, such an integration necessitates token-level BM25 guidance. This insight leads to our core contribution: \textbf{relevance-guided attention}---a mechanism that biases attention computations using explicit relevance signals. 

We instantiate this idea in \textsc{REGENT} (RElevance-Guided ENTity-aware reranker).\footnote{\textbf{Code and data available at:https://github.com/shubham526/SIGIR-AP-2025-REGENT}} \textsc{REGENT} is built on a multi-vector architecture designed to process two distinct types of information in parallel: one pathway handles the fine-grained vectors representing document tokens, while a second pathway processes the high-level vectors representing semantic entities. It then guides attention using two complementary signals: token-level BM25 scores to highlight lexically relevant terms and query-specific entity representations to focus on semantically important concepts. 

A multi-vector architecture naturally supports this dual guidance, preserving fine-grained token interactions while enabling coordinated entity-level processing. When processing the dot-com query, \textsc{REGENT} learns to prioritize passages mentioning relevant entities like ``\texttt{Nasdaq}'' and ``\texttt{venture capital}'', while the BM25 signals simultaneously highlight exact matches such as \texttt{bubble} and \texttt{crash}. These lexical and semantic signals work together to direct attention toward the most relevant content, even when it is scattered across a long document. We show that \textsc{REGENT} achieves significant improvements across three large-scale datasets featuring long-form news and social science articles paired with complex information needs. The model outperforms strong neural baselines including recent LLM-based re-rankers such as RankVicuna~\cite{pradeep2023rankvicuna} and RankZephyr~\cite{pradeep2023rankzephyr} with gains of up to 108\% over BM25. Crucially, removing the model's entity component--which captures relationships between key entities--results in a 74\% drop in performance, underscoring the essential role of the semantic skeleton in effective long-document retrieval.

\paragraph{\textbf{Contributions.}} We make three key contributions to neural IR: (1) \textbf{Relevance-guided attention:} the first mechanism to directly weave lexical and entity signals into attention computations, shifting from post-hoc score combination to integrated neural guidance during ranking; (2) \textbf{Token-level BM25 integration:} a principled method for incorporating fine-grained lexical evidence into transformer attention, enabling models to focus on individually relevant terms rather than coarse document-level signals; and (3) \textbf{Dynamic entity-aware processing:} a dual-pathway architecture that learns query-specific entity relationships through dedicated attention layers, moving beyond static similarity matrices to context-sensitive semantic reasoning.

The remainder of this paper is structured as follows. We first review related work in entity-oriented search and neural IR in Section~\ref{sec:Related Work}. We then detail the \textsc{REGENT} architecture, including its relevance-guided attention mechanism, in Section\ref{sec:Approach}. Our experimental setup, datasets, and baselines are described in Section\ref{sec:Experimental Methodology}. We present and analyze our results in Section~\ref{sec:Results}, including a series of ablation studies and qualitative analyses. Finally, we conclude our work in Section~\ref{sec:Conclusion}.

\section{Related Work}
\label{sec:Related Work}

\textbf{Entity-Oriented Search.}  Early work such as EQFE~\cite{dalton2014entity} leveraged entity-based features, demonstrating the utility of structured signals. This evolved into latent semantic projection approaches~\cite{gabrilovich2009wikipedia,liu2015latent}, where queries and documents were mapped to an entity space to capture hidden semantic relationships. The move toward explicit entity-term integration began with early entity-based language models~\cite{raviv2016document,ensan2017document}, which balanced term- and entity-level signals. The Word-Entity Duet~\cite{xiong2017word} introduced rich four-way interactions between queries and documents, while Explicit Semantic Ranking~\cite{xiong2017explicit} employed knowledge graphs for structured matching. Approaches such as EDRM \cite{liu-etal-2018-entity} incorporate knowledge graph semantics into neural ranking models through pre-computed entity similarities.

\textbf{Neural Information Retrieval.} The evolution of neural IR can be traced through several paradigmatic shifts. Pre-BERT approaches followed two main trajectories: representation-based models that used static embeddings~\cite{huang2013learning,shen2015entity,mitra2019an} and interaction-based models that computed term-level similarity matrices~\cite{guo2016drmm,xiong2017end,hui-etal-2017-pacrr}. The introduction of BERT~\cite{devlin2018bert} marked a transformative moment, spawning innovations like cross-encoders for fine-grained query-document interaction~\cite{akkalyoncu-yilmaz-etal-2019-cross,dai2019deeper} and bi-encoders such as DPR~\cite{karpukhin-etal-2020-dense} and ANCE~\cite{xiong2020ance} that balanced effectiveness with efficiency.

Recent advances have moved beyond single-vector representations to capture richer query-document interactions. In this regard, ColBERT~\cite{khattab2020colbert} pioneered ``late interaction'' mechanisms using dense token-level representations while poly-encoders~\cite{humeau2020polyencoders} generate multiple document representations through learned context codes. %ME-BERT~\cite{luan-etal-2021-sparse} produces multiple representations from document prefixes, enabling efficient nearest neighbor search. %These multi-vector approaches preserve fine-grained matching capabilities while maintaining computational tractability.

Neural approaches have also transformed pseudo-relevance feedback by leveraging embedding-based similarity measures. 
%Neural PRF~\cite{li-etal-2018-nprf} and BERT-QE~\cite{zheng-etal-2020-bert} employ neural ranking models to assess similarity between documents and top-ranked feedback documents, moving beyond traditional statistical approaches. 
While ANCE-PRF~\cite{hongchien2021ance-prf} enhances retrieval effectiveness by retraining the query encoder using PRF information, ColBERT-PRF~\cite{xiao2023colbert-prf} directly utilizes BERT embeddings for retrieval without further training, avoiding topic drift issues common with polysemous words.

\textbf{Hybrid Retrieval.} Prior work has also shown how to blend traditional (sparse) IR signals with transformers. DeepCT~\cite{dai2020context}, docT5query~\cite{nogueira2019document}, and CEQE~\cite{naseri2021ceqe} identify important terms or expansions using contextualized models.  While the original DPR paper~\cite{karpukhin-etal-2020-dense} reported limited gains from combining dense and sparse signals, subsequent work has shown that hybrid approaches can consistently outperform purely dense or sparse models. \citet{ma2021a} demonstrated that carefully designed dense-sparse hybrids yield significant improvements across multiple benchmarks. \texttt{CLEAR}~\cite{gao2021complement} proposed a jointly trained hybrid where the dense encoder complements BM25 by capturing semantic matches it misses. TCT-ColBERT~\cite{lin-etal-2021-batch} used knowledge distillation with ColBERT as a teacher to produce soft labels on the fly, combining these with BM25 and doc2query-T5 expansion for strong effectiveness with high efficiency. Askari et al.~\cite{askari2023injecting} showed that injecting BM25 scores as special tokens into transformer inputs improves re-ranking, offering direct evidence that lexical signals can be leveraged within neural models. Recent advancements include SPLADE-v3~\cite{lassance2024spladev3}, which improves lexical expansion with better regularization, and LexMAE~\cite{shen2023lexmae}, which introduces pre-training objectives that preserve lexical sensitivity. Meanwhile, hybrid systems have begun incorporating LLMs, as seen in HyDE~\cite{gao-etal-2023-precise} and Promptagator~\cite{dai2022promptagator}, which use hypothesized documents or synthetic queries to improve retrieval.

\textbf{LLM-based Re-Rankers.} 
LLMs have recently shown strong zero-shot reranking performance. RankVicuna~\cite{pradeep2023rankvicuna} achieved near-GPT-3.5 results on TREC DL using a 7B model. RankZephyr~\cite{pradeep2023rankzephyr} further closed the gap with GPT-4 while demonstrating robust generalization across BEIR and NovelEval. More advanced methods such as RankR1~\cite{zhuang2025rankr1} introduced reasoning via reinforcement learning, yielding large gains on complex and out-of-domain queries with minimal training data. Search-R1~\cite{jin2025searchr1} extended this by teaching LLMs to iteratively generate and refine search queries during reasoning, improving performance on retrieval-augmented QA tasks.

%The development of open-source LLM-based re-rankers provides important context for large-scale ranking architectures. RankVicuna \cite{pradeep2023rankvicuna} demonstrated high-quality zero-shot listwise reranking using a 7B model, approaching GPT-3.5 performance on TREC DL benchmarks. RankZephyr \cite{pradeep2023rankzephyr} further advanced the field by narrowing the gap with GPT-4 while exhibiting strong generalization across diverse datasets including BEIR and NovelEval. More sophisticated reasoning approaches have emerged through RankR1 \cite{zhuang2025rankr1}, which introduces reasoning into reranking through reinforcement learning, achieving significant improvements especially on complex and out-of-domain queries while using only a fraction of the training data. Search-R1 \cite{jin2025searchr1} extends this paradigm by teaching LLMs to iteratively issue and refine search queries during reasoning, yielding substantial gains in retrieval-augmented QA.

\textbf{Advances in Training and Representation.}  %Recent progress in dense retrieval pre-training offers key insights for integrating diverse signal types. 
RetroMAE~\cite{xiao-etal-2022-retromae} introduced retrieval-specific masked autoencoding, outperforming general language modeling by aligning training objectives with retrieval tasks. E5~\cite{wang2024textembeddings} extended this with weakly supervised contrastive learning, achieving strong zero-shot performance and demonstrating the effectiveness of large-scale contrastive pre-training for general-purpose embeddings. Instruction-based models such as INSTRUCTOR~\cite{su-etal-2023-one} added architectural flexibility by generating embeddings conditioned on natural language prompts, thus enabling a single model to adapt across retrieval scenarios without task-specific fine-tuning.

\section{Approach}
\label{sec:Approach}

Given a complex query $Q$ and a set of long candidate documents $\mathcal{D}$, our goal is to \textit{re-rank} the candidates by their relevance to $Q$. We propose \textsc{REGENT}, a re-ranking model built around a novel \textbf{relevance-guided attention mechanism}. This is our key innovation. Unlike standard attention, which relies only on learned token interactions, our approach explicitly guides attention using two relevance signals: (1) \textbf{token-level BM25 scores} that enhance key and value representations for lexically matched terms, and (2) \textbf{query-specific entity representations} that modulate attention weights based on semantic relevance. This allows the model to integrate lexical and semantic signals directly during attention computation, rather than fusing them post hoc, as done by prior work~\cite{askari2023injecting,wang2021bert}

%Unlike standard attention that treats all token interactions uniformly, our approach biases attention computations using explicit relevance signals: (1) \textbf{token-level BM25 scores} that directly enhance key and value representations for lexically matching terms, and (2) \textbf{dynamic entity context} that influences token attention weights based on surrounding entity semantics. This enables the model to simultaneously leverage fine-grained lexical evidence and high-level semantic structure during attention computation, rather than combining these signals post-hoc.

 % Given a complex query $Q$ and a set of long candidate documents $\mathcal{D}$, our goal is to \textit{re-rank} the candidates by their relevance to $Q$. We propose \texttt{REGENT}, a multi-vector neural re-ranking model that leverages two key innovations: (1) token-level BM25 integration to directly influence term representations and (2) entity-aware attention to capture contextual relationships. Instead of treating BM25 scores and entity information as independent signals, our approach enables their interaction within the attention mechanism, allowing the model to dynamically adjust term importance based on both lexical matching evidence and entity-level context. %We use our approach to re-rank a candidate set of 1000 documents per query obtained using BM25 (\( k_1 = 1.2, b = 0.75 \)). 

\subsection{Model Architecture}
\label{subsec:Model Architecture}

%\textbf{Model Architecture.} 
% At its core, \texttt{REGENT} builds upon a pre-trained BERT \cite{devlin2018bert} encoder to generate initial contextual representations for both queries and documents. Given a query-document pair, we first obtain their respective token embeddings through BERT (\texttt{bert-base-uncased}). Input sequences are truncated or padded to 512 tokens for consistent processing. These embeddings then pass through multiple layers of our enhanced cross-attention mechanism, where each layer consists of an entity-aware attention module followed by a feed-forward network. The entity-aware attention takes the query embeddings as queries and document embeddings as keys/values, incorporating both BM25 signals and entity information (detailed below). After processing through $L$ attention layers, we apply mean pooling over the final query representation to obtain a fixed-size vector. We stack $L=2$ attention layers, each with eight attention heads, to capture rich token-level interactions. This pooled representation is then processed through a sophisticated scoring network consisting of multiple feed-forward layers with residual connections. Specifically, each layer applies \texttt{GELU} activation and \texttt{LayerNorm}, progressively reducing the dimension while maintaining normalized activations. The final single-dimensional output represents the relevance score.
At its core, \textsc{REGENT} uses a pre-trained \texttt{bert-base-uncased} encoder~\cite{devlin2018bert} to encode queries and documents separately. Inputs are padded or truncated to 512 tokens.
These embeddings are passed through two cross-attention layers (each with eight heads), where queries attend to document tokens \emph{via an entity-aware attention module}, followed by a feed-forward network. This module integrates BM25 scores and entity signals to guide attention. The final query representation is mean-pooled and fed into a multi-layer scoring network with residual connections, \texttt{GELU} activations, and \texttt{LayerNorm}, producing a single relevance score via a linear output layer.

\subsection{Token-Level BM25 Integration}
\label{subsec:Token-Level BM25 Integration}

%\textbf{Token-Level BM25 Integration.} 
\textbf{BM25 Score Alignment with Subword Tokens.} Given a query-document pair, we compute a ranking signal vector $r \in \mathbb{R}^n$ where each element $r_i$ represents the BM25 score for the $i$-th token position. These BM25 scores are computed using Lucene's EnglishAnalyzer, which applies standard IR preprocessing including stopword removal, stemming, and lowercasing. To bridge the gap between this preprocessing and BERT's tokenization, we track how each word splits into subword tokens. During indexing, we store a tuple \( (w_i, s_i, e_i) \) where \( w_i \) is the word index, and \( s_i, e_i \) denote the start and end positions of its subword tokens. To obtain token-level BM25 scores, we first compute term-level scores and then propagate them to all corresponding subword tokens. For example, both ``play'' and ``\#\#ing'' inherit the BM25 score of ``playing''.

\smallskip
\noindent \textbf{Why Not Tokenize BM25 Terms Directly?} We considered an alternative approach of re-tokenizing Lucene's preprocessed (e.g., stemmed) terms with the BERT tokenizer and assigning BM25 scores directly to the resulting subword units. However, this strategy is conceptually problematic because BM25 is defined over full words or terms, not arbitrary subword pieces like ``\#\#ing'' or ``\#\#ly,'' which lack independent frequency statistics or meaningful retrieval signals. Therefore, propagating word-level scores across subword spans, as done in our approach, is a more principled and interpretable integration of BM25 into subword-based neural architectures. This approach ensures that stopwords and punctuation retained by BERT (e.g., ``a'', ``the'', ``of'') receive minimal or zero BM25 scores since they were removed or heavily downweighted in the Lucene preprocessing pipeline. Consequently, these tokens have negligible impact on the attention mechanism despite being present in BERT's representation.

\smallskip
\noindent \textbf{Integrating BM25 into Attention via Key and Value Enhancement.} We use these token-level scores to improve key representations ($K$) and value ($V$) representations in the attention mechanism. In a standard attention mechanism, Key ($K$) and Value ($V$) matrices are learned projections of the input token embeddings. Our approach enhances these matrices to make lexically important tokens stand out more, influencing both which tokens are attended to (via keys) and how much they contribute to the output (via values). Specifically, let $R \in \mathbb{R}^{n \times d}$ be the matrix formed by repeating the BM25 score vector $r$ for each dimension of the embedding space, where $d$ is the embedding dimension. We then compute $K' = K + \alpha \cdot R$ and $V' = V + \alpha \cdot R$. Here, $K, K' \in \mathbb{R}^{n \times d}$ are the original and enhanced key matrices, $V, V' \in \mathbb{R}^{n \times d}$ are the original and enhanced value matrices, and $\alpha \in \mathbb{R}$ is a learnable scalar parameter that controls the influence of BM25 scores on the attention mechanism. Our intuition is that this enhancement would bias key representations toward strong lexical matches and enriches value representations to emphasize high-BM25 tokens, while minimizing the influence of less informative tokens. By influencing both attention weights (via keys) and contextual representations (via values), the model would be able to learn matching patterns that effectively integrate traditional lexical signals with neural semantic evidence.

\subsection{Query-Specific Entity Set Construction}
\label{subsec:Query-Specific Entity Set Construction}

%\textbf{Query-Specific Entity Set Construction.}
As discussed in Section~\ref{sec:Introduction}, entities can provide a meaningful semantic scaffold to organize document content. However, using all entities indiscriminately can introduce noise and dilute the focus of the attention mechanism.
%For example, a query about ``\textit{inflation impact on trade}'' needs only a small subset of the entities in an economics article. 
To address this, we design a multi-stage pipeline that builds a focused \emph{query-specific} set of entity representations. 
%This ensures that only contextually relevant entities guide REGENT’s attention, avoiding the limitations of static, query-agnostic approaches used in prior work.
%
First, we retrieve the top 1,000 documents using BM25 and aggregate all unique entities linked within these documents to form a candidate pool. Next, we train a separate BERT-based entity ranking model (via 5-fold CV) to score each candidate entity's relevance to the query. This ranker is a standard BERT cross-encoder. For each query-entity pair, we format the input as ``\texttt{[CLS]} query text \texttt{[SEP]} entity name \texttt{[SEP]}''. The final relevance score is derived from the \texttt{[CLS]} token's output embedding, which is passed through a linear layer. This ranker is trained once and reused at inference time to efficiently score entities given a new query. Since the candidate pool remains fixed, we can precompute document-entity mappings and quickly compute query-document relevance scores using only the top-ranked entities. 

To train the ranker, we follow established practice~\cite{nguyen2016marco,dietz2017trec,dietz2018trec}, treating entities found in documents marked relevant in the qrels as positive examples and all others as negative. Using the trained ranker, we then score all candidate entities for each query and retain the top 20 as the query-relevant entity set. For each query-document pair, we take the intersection of the document's entities with this set and scale each entity's pre-trained Wikipedia2Vec \cite{yamada-etal-2020-wikipedia2vec} embedding by its corresponding ranker score. This results in a \emph{query-specific} set of entity embeddings that \textsc{REGENT}'s entity-aware attention mechanism (described below) can use to focus on semantically meaningful signals.

\subsection{Entity-Aware Dual-Pathway Attention}
\label{subsec:Entity-Aware Dual-Pathway Attention}

%\textbf{Entity-Aware Dual-Pathway Attention.} 
We employ parallel token and entity attention pathways to enhance relevance modeling. The token pathway captures lexical matches using BM25 signals:
\begin{equation}
A_t = \text{softmax}\left(\frac{QK'^T}{\sqrt{d_k}}\right)V'
\end{equation}
where $Q$ represents query token embeddings, $K'$, $V'$ are the BM25-enhanced document representations, and $d_k$ is the scaling factor.

The entity pathway processes the refined entity sets generated by the pipeline described in Section \ref{subsec:Query-Specific Entity Set Construction}. It establishes semantic context in two steps: First, it computes attention between the query's entity set and the document's (filtered) entity set to identify relevant entity matches:
\begin{equation}
A_e = \text{softmax}\left(\frac{E_q W_q^e (E_d W_k^e)^T}{\sqrt{d_k}}\right)E_d W_v^e
\end{equation}
where $E_q \in \mathbb{R}^{n_q \times d_e}$ and $E_d \in \mathbb{R}^{n_d \times d_e}$ are query and document entity embeddings, respectively, which have been prescaled by their relevance scores as determined by our entity ranking pipeline. $W_q^e$, $W_k^e$, $W_v^e$ are learnable query, key, and value projections specific to entity-entity attention. We implement this using multi-head attention to capture different types of entity relationships.

Then, it allows these entity matches to influence token representations through a separate attention mechanism:
\begin{equation}
A_{et} = \text{softmax}\left(\frac{QW_q^t(A_eW_k^t)^T}{\sqrt{d_k}}\right)A_eW_v^t
\end{equation}
where $Q$ represents query token embeddings, and $W_q^t$, $W_k^t$, $W_v^t$ are learnable projections specifically for entity-token interactions. Importantly, the projection matrices used in equations (2) and (3) are \textit{not} shared ($W_q^e \neq W_q^t$, $W_k^e \neq W_k^t$, $W_v^e \neq W_v^t$) because they serve fundamentally different purposes: the former transforms entity embeddings for entity-entity matching, while the latter projects entity attention outputs for token-level contextualization. This allows the model to learn distinct transformation patterns for each attention mechanism: one optimized for entity relationship detection and another for incorporating entity context into token representations.

Our intuition is that this two-step process enables context-aware token matching. First, $A_e$ aligns related entities across the query and document (e.g., linking ``\texttt{Nasdaq}'' in the query to ``\texttt{tech stocks}'' in the document). Then, $A_{et}$ uses this entity context to guide token attention ---- helping the model prioritize mentions of ``crash'' or ``valuation'' that appear near relevant entities like ``\texttt{Amazon}'' or ``\texttt{venture capital}'', rather than in unrelated sections.

% Our intuition is that this two-step process enables contextually-aware token matching. In $A_e$, learned entity projections allow the model to identify semantically related entities between query and document (e.g., matching symptom entities like \texttt{fatigue} to similar health-related entities in the query). Then, in $A_{et}$, this entity-level semantic context influences token matching---when looking for terms like ``vitamin D'', the model can distinguish between mentions that appear in relevant medical contexts (near symptom entities) versus less relevant contexts (near dietary supplement entities).

\subsection{Adaptive Fusion Mechanism}
\label{subsec:Adaptive Fusion Mechanism}

%\textbf{Adaptive Fusion Mechanism}
The token and entity pathways are combined through a fusion mechanism. First, we compute attention weights: 
\begin{equation}
\alpha = \sigma(\text{LayerNorm}(W_f[A_t; A_{et}]))
\end{equation}
where $\sigma$ is the sigmoid function, $W_f$ is a learnable projection, and $[;]$ denotes concatenation. The final output is then computed as:
\begin{equation}
O = \alpha \odot A_t + (1 - \alpha) \odot A_{et}
\end{equation}

Although both tokens and entities can have BM25 scores, we integrate them through separate pathways to reflect their distinct roles: tokens capture lexical matches, while entities represent higher-level semantic concepts. Token scores reflect term frequency, while entity relevance depends on contextual importance. Our dual-pathway design leverages this distinction---lexical cues guide token attention, while entity context informs semantic focus. An adaptive fusion layer balances the two; our experiments show that this separation outperforms unified models.

\smallskip
\section{Experimental Setup}
\label{sec:Experimental Methodology}

\subsection{Datasets}
\label{subsec:Datasets}

We evaluate \textsc{REGENT} on three benchmarks selected for their (1) long-form documents with high entity density and (2) complex, non-factual information needs requiring semantic reasoning. \textbf{CODEC}~\cite{mackie2022codec} features 42 social science queries over 729K entity-linked documents (avg. 159 entities per document), with 6,186 expert-annotated relevance judgments. \textbf{TREC Robust 2004}~\cite{voorhees2003overview} includes 250 challenging queries designed to stress lexical models, covering 528K articles (avg. 116 entities per document) and 311,409 graded relevance labels. We specifically use the ``title'' queries, which is a standard and challenging sub-task for this collection \textbf{TREC Core 2018}~\cite{allan2017trec} contains 50 complex queries over 595K news/blog posts (avg. 123 entities per document) with 26,233 graded judgments. 

We deliberately exclude benchmarks like MS MARCO and TREC DL, as their focus on short passages and factoid queries with minimal entity presence (e.g., MS MARCO averages 1.7–2.1 entities per document, and over 97\% of its queries mention at most one entity~\cite{kamphuis2023}) makes them fundamentally ill-suited for evaluating the core contribution of \textsc{REGENT}, which is its ability to reason over rich entity relationships in long documents. Our selected datasets follow established practice in entity-oriented IR~\cite{dalton2014entity,nguyen2024dyvo} and better capture the challenges of long-document retrieval

%We select benchmarks that meet two key criteria: (1) long-form documents with high entity density, and (2) complex, non-factual information needs requiring semantic reasoning. Based on these criteria, we evaluate \textsc{REGENT} on three established benchmarks: (1) \textbf{CODEC}~\cite{mackie2022codec} consists of 42 social science queries over 729K entity-linked documents, with an average of 159 entities per document. It includes 6,186 expert-annotated relevance judgments derived from both manual and automated retrieval runs. (2) \textbf{TREC Robust 2004}~\cite{voorhees2003overview} features 250 challenging queries designed to stress lexical models, evaluated over 528K articles (average 116 entities per document) from TREC disks 4 and 5 (excluding the Congressional Record), with 311,409 graded relevance judgments. (3) \textbf{TREC Core 2018}~\cite{allan2017trec} includes 50 complex queries over 595K news and blog posts from the TREC Washington Post v2 collection (average 123 entities per document), with 26,233 graded relevance labels.

\smallskip
\noindent  \textbf{\textit{Evaluation.}} We report Prec@20, nDCG@20, and Mean Average Precision (MAP) using the official \texttt{trec\_eval} tool (using \texttt{-c} flag). 

%For \textbf{evaluation}, we report Prec@20, nDCG@20, and MAP using the official \texttt{trec\_eval} tool (using \texttt{-c} flag). 

% \smallskip
% \noindent \textbf{\textit{A Note on Benchmark Selection.}} We exclude popular benchmarks like MS MARCO and TREC DL, as their focus on short passages and factoid queries makes them ill-suited for evaluating REGENT. MS MARCO, for example, contains few entities (avg 1.7-2.1 per document) ~\cite{kamphuis2023}. Moreover, our preliminary analysis shows that more than 97\% of the MS MARCO queries mention one or zero entities, limiting its utility to evaluate entity relationship modeling. In contrast, our selected datasets follow established practice in entity-oriented IR~\cite{dalton2014entity,nguyen2024dyvo} and better reflect the challenges of long-document retrieval and complex semantic reasoning.

\subsection{Baselines}
\label{subsec:Baselines}

% To validate the effectiveness of \textsc{REGENT}, we compare it against a comprehensive suite of state-of-the-art models representing the major paradigms in neural re-ranking. All models that require fine-tuning were trained on our target datasets using the same 5-fold cross-validation setup as REGENT to ensure a fair and rigorous comparison. Our baselines fall into five main categories:
To validate \textsc{REGENT}, we compare it against a broad set of state-of-the-art neural re-ranking models. All fine-tuned models are trained on the target datasets using the same 5-fold cross-validation as \textsc{REGENT} for fair comparison. Baselines span five main categories:

\textbf{1. Cross-Encoder Models.}
These models represent the standard and powerful paradigm of fine-tuning a transformer for pointwise re-ranking by processing a concatenated \texttt{[CLS] query [SEP] document} sequence. We include a wide range of established and recent architectures: \textbf{BERT}~\cite{devlin2018bert}, \textbf{RoBERTa}~\cite{liu2019roberta}, \textbf{DeBERTa}~\cite{he2020deberta}, \textbf{ELECTRA}~\cite{clark2020electra}, \textbf{ConvBERT}~\cite{jiang2020convbert}, and the sequence-to-sequence \textbf{RankT5}~\cite{zhuang2023rankt5}. We also include several zero-shot cross-encoders from the Sentence-BERT library~\cite{reimers2019sentence}.

\textbf{2. Bi-Encoder and Multi-Vector Models.}
%To compare against models that also use disaggregated representations, we include several strong baselines. 
We include a fine-tuned \textbf{ColBERT v2}~\cite{khattab2020colbert} model. We also test a suite of bi-encoder models, including a zero-shot version of \textbf{DPR}~\cite{karpukhin-etal-2020-dense} and various pre-trained models from the Sentence-BERT library to represent the efficient dual-encoder paradigm.

\textbf{3. Entity-Aware Models.}
To compare specifically with other methods that leverage entity knowledge, we include two key baselines. \textbf{EDRM}~\cite{liu-etal-2018-entity} is a pioneering neural model that uses pre-computed entity similarity matrices. \textbf{ERNIE}~\cite{zhang-etal-2019-ernie} is a knowledge-enhanced pre-trained model that integrates entity information directly during its pre-training phase. These models allow us to assess the benefits of \textsc{REGENT}'s dynamic, relevance-guided entity processing against more static approaches.

\textbf{4. LLM-Based Re-Rankers.}
To situate \textsc{REGENT}'s performance against the latest LLM-based rerankers, we include several powerful, publicly available zero-shot re-rankers. This includes the pointwise \textbf{BGE-ReRanker-v2-Gemma}~\cite{bge_m3}, \textbf{INSTRUCTOR}~\cite{su-etal-2023-one}, and the listwise \textbf{RankVicuna}~\cite{pradeep2023rankvicuna} and \textbf{RankZephyr}~\cite{pradeep2023rankzephyr}, which have demonstrated performance competitive with proprietary models.

\textbf{5. Hybrid Models.}
Finally, to contrast \textsc{REGENT}'s deep signal fusion with other hybrid approaches, we include several models that also combine lexical and semantic signals. We re-implement the work by ~\citet{askari2023injecting} that injects BM25 scores directly into the input text sequence. We also re-implement work by ~\citet{wang2021bert} that linearly interpolates the final scores from a dense model and BM25. Lastly, we include several strong \textbf{Coordinate Ascent} baselines, which iteratively optimize the weights for combining BM25 with scores from various fine-tuned and zero-shot models. 

\smallskip
\textbf{\textit{Additional Baselines on TREC Robust 2004.}}
Given the historical significance of the TREC Robust 2004 track, we include several additional influential full retrieval (1st stage, not reranking) models to provide a broader context for our results on this specific dataset. The results are shown in Table~\ref{tab:res-tab-full-ret}.

\smallskip 
\textbf{Note.} We acknowledge that the field of LLM-based re-ranking is evolving rapidly; for instance, RankR1~\cite{zhuang2025rankr1} and Search-R1~\cite{jin2025searchr1} were published concurrently with our work. Our selection of baselines therefore focuses on a representative and diverse set of strong, publicly available models at the time of our experiments.

\subsection{Implementation Details}
\label{subsec:Implementation Details}

\noindent \textbf{Model Details.} We implement our model in PyTorch using the HuggingFace library with \texttt{bert-base-uncased} as the base encoder (parameters not frozen). We fine-tune the model using binary cross-entropy loss with Adam~\cite{kingma2014adam} optimizer at a learning rate of $2\text{e}{-5}$ and apply a linear warmup over the first 1{,}000 steps. Training is performed for 10 epochs with a batch size of 8, and early stopping based on validation MAP (computed via \texttt{pytrec\_eval}). We apply gradient clipping and a dropout rate of 0.1 to prevent overfitting. Document embeddings and entity links are precomputed and cached for efficient inference. For all re-ranking experiments, the candidate set consists of the top 1000 documents retrieved by the initial BM25 ranking. This same set was provided to \textsc{REGENT} and all re-ranking baselines to ensure a fair comparison.

%All experiments were run on a single NVIDIA RTX 3090 (24GB).

% \noindent \textbf{Model Details.} We implemented our model using PyTorch and the HuggingFace library.  We used the \texttt{bert-base-uncased} model as the base encoder with a maximum sequence length of 512 tokens. The model was fine-tuned using the \texttt{CrossEntropyLoss} function provided by PyTorch. BERT parameters are not frozen during training. 
% %
% For optimization, we employed the Adam optimizer \cite{kingma2014adam} with a learning rate of $2e-5$ and a linear warmup schedule over the first 1,000 steps. The training process used a batch size of 8 and was carried out for 10 epochs, with the model evaluated after each epoch. Early stopping was implemented based on the validation MAP score, calculated using \texttt{pytrec\_eval}, and the best-performing checkpoint was saved. To prevent overfitting, we apply gradient clipping and a dropout rate of 0.1.
% %
% To enhance inference efficiency, document embeddings and entity links were precomputed and cached offline. All experiments were conducted on a single NVIDIA RTX 3090 GPU with 24 GB of memory.

\smallskip

%\noindent \textbf{Train and Test Data.} Positive training examples are documents labeled as relevant in the dataset. Following \citet{karpukhin-etal-2020-dense}, negatives are sampled from a BM25 candidate ranking (Pyserini default) if explicitly marked irrelevant or unjudged. We balance positives and negatives and perform 5-fold cross-validation at the query level.

\noindent \textbf{Train and Test Data.}  As positive examples during training, we use documents that are assessed as relevant in the ground truth provided with the dataset. Following the standard \cite{karpukhin-etal-2020-dense}, for negative examples, we use documents from a BM25 candidate ranking (obtained using Pyserini default) which are either explicitly annotated as negative or not present in the ground truth. We balance the training data by keeping the number of negative examples the same as the number of positive examples. These examples are then divided into 5-folds for cross-validation. We create these folds at the query level.

%\smallskip

%\noindent \textbf{Candidate Ranking.} We retrieve a candidate set of 1000 documents per query using \texttt{BM25+RM3} (Pyserini default). 

%\smallskip

% \noindent \textbf{Evaluation Metrics.} (1) Precision at $k=20$, (2) Normalized Discounted Cumulative Gain (nDCG) at $k=20$, and (3) Mean Average Precision (MAP). We use the official \texttt{trec\_eval} tool from NIST (with the \texttt{-c} flag) to evaluate each system.  We conduct significance testing using paired-t-tests. 

\smallskip

\noindent \textbf{Entity Linking and Embeddings.} We use pre-trained Wikipedia2Vec embeddings~\cite{yamada-etal-2020-wikipedia2vec} from \citet{Kamphuis2023mmead} as base entity representations and generate initial links using WAT~\cite{piccinno2014wat}, an open-source, deterministic EL system. We prefer WAT over LLM-based methods to ensure reproducibility, scalability, and avoid proprietary/API constraints. Recent work~\cite{xin2024,ding2024,vollmers-etal-2025-contextual} also shows that traditional EL systems outperform LLMs, especially for rare or ambiguous entities that LLMs often misidentify. As described in Section~\ref{subsec:Query-Specific Entity Set Construction}, these initial links are refined through an entity selection and classification pipeline to produce a query-specific, relevance-weighted entity set for \textsc{REGENT}. Although we use standard tools, \textsc{REGENT} is EL-agnostic: its architecture does not rely on any specific linking method, allowing future EL improvements to be easily incorporated.

\section{Results and Analysis}
\label{sec:Results}

\begin{table*}[t]
\centering
\caption{Results for re-ranking on Robust04 (Title), Core18, and CODEC datasets.
Statistical significance is determined using paired t-tests ($p<0.05$) with Bonferroni correction ($\blacktriangle$ indicates significantly better, $\blacktriangledown$ indicates significantly worse than BM25).}
\label{tab:results}
\scalebox{0.82}{
\begin{tabular}{
  @{}l
  S[table-format=1.3, table-space-text-post={$\blacktriangle$}]
  S[table-format=1.3, table-space-text-post={$\blacktriangle$}]
  S[table-format=1.3, table-space-text-post={$\blacktriangle$}]
  S[table-format=1.3, table-space-text-post={$\blacktriangle$}]
  S[table-format=1.3, table-space-text-post={$\blacktriangle$}]
  S[table-format=1.3, table-space-text-post={$\blacktriangle$}]
  S[table-format=1.3, table-space-text-post={$\blacktriangle$}]
  S[table-format=1.3, table-space-text-post={$\blacktriangle$}]
  S[table-format=1.3, table-space-text-post={$\blacktriangle$}]@{}
}
\toprule
& \multicolumn{3}{c}{\textbf{TREC Robust04 (Title)}} 
& \multicolumn{3}{c}{\textbf{TREC Core 2018}} 
& \multicolumn{3}{c}{\textbf{CODEC}} \\
\midrule
& \textbf{MAP} & \textbf{nDCG@20} & \textbf{P@20} 
& \textbf{MAP} & \textbf{nDCG@20} & \textbf{P@20}
& \textbf{MAP} & \textbf{nDCG@20} & \textbf{P@20} \\
\midrule
BM25 & 0.292 & 0.435 & 0.384 & 0.315 & 0.447 & 0.459 & 0.363 & 0.380 & 0.432 \\
\midrule
\multicolumn{10}{l}{\textit{\textbf{Cross-Encoder Models (Fine-Tuned)}}} \\
RankT5 \cite{zhuang2023rankt5} & 0.303 & 0.494$\blacktriangle$ & 0.429$\blacktriangle$ & 0.216$\blacktriangledown$ & 0.326$\blacktriangledown$ & 0.334$\blacktriangledown$ & 0.380 & 0.437 & 0.452$\blacktriangle$ \\
MonoBERT \cite{nogueira2019passage} & 0.297 & 0.479 & 0.409 & 0.302 & 0.469 & 0.461 & 0.382 & 0.440 & 0.463$\blacktriangle$ \\
RoBERTa \cite{liu2019roberta} & 0.290 & 0.474 & 0.410 & 0.260$\blacktriangledown$ & 0.361$\blacktriangledown$ & 0.388$\blacktriangledown$ & 0.359$\blacktriangledown$ & 0.377$\blacktriangledown$ & 0.433$\blacktriangle$ \\
DeBERTa \cite{he2020deberta} & 0.293 & 0.486 & 0.422 & 0.346 & 0.519$\blacktriangle$ & 0.507 & 0.390 & 0.449 & 0.463$\blacktriangle$ \\
ELECTRA \cite{clark2020electra} & 0.268 & 0.446 & 0.387 & 0.240$\blacktriangledown$ & 0.353 & 0.361 & 0.323$\blacktriangledown$ & 0.317$\blacktriangledown$ & 0.385$\blacktriangledown$ \\
ConvBERT \cite{jiang2020convbert} & 0.321$\blacktriangle$ & 0.519$\blacktriangle$ & 0.451$\blacktriangle$ & 0.321 & 0.496 & 0.489 & 0.382 & 0.441 & 0.462$\blacktriangle$ \\
ERNIE \cite{zhang-etal-2019-ernie} & 0.289 & 0.475 & 0.412 & 0.340 & 0.520 & 0.507 & 0.391 & \textbf{0.457} & 0.474$\blacktriangle$ \\
EDRM \cite{liu-etal-2018-entity} & 0.067$\blacktriangledown$ & 0.102$\blacktriangledown$ & 0.094$\blacktriangledown$ & 0.092$\blacktriangledown$ & 0.098$\blacktriangledown$ & 0.134$\blacktriangledown$ & 0.298$\blacktriangledown$ & 0.289$\blacktriangledown$ & 0.352$\blacktriangledown$ \\
\midrule
\multicolumn{10}{l}{\textit{\textbf{SentenceBERT Cross-Encoders (Zero-Shot)}}} \\
ms-marco-MiniLM-L6-v2 \cite{reimers2019sentence} & 0.275 & 0.447 & 0.392 & \multicolumn{1}{c}{0.272} & \multicolumn{1}{c}{0.434} & \multicolumn{1}{c}{0.429} & \multicolumn{1}{c}{0.365} & \multicolumn{1}{c}{0.426} & \multicolumn{1}{c}{0.436} \\
ms-marco-electra-base \cite{reimers2019sentence} & 0.233$\blacktriangledown$ & 0.405 & 0.346 & 0.175$\blacktriangledown$ & 0.289$\blacktriangledown$ & 0.306$\blacktriangledown$ & \multicolumn{1}{c}{0.367} & \multicolumn{1}{c}{0.423} & \multicolumn{1}{c}{0.447} \\
qnli-distilroberta-base \cite{reimers2019sentence} & 
 0.067$\blacktriangledown$ & 0.105$\blacktriangledown$ & 0.103$\blacktriangledown$ & 0.047$\blacktriangledown$ & 0.036$\blacktriangledown$ & 0.046$\blacktriangledown$ & 0.298$\blacktriangledown$ & 0.294$\blacktriangledown$ & 0.347$\blacktriangledown$ \\
quora-distilroberta-base \cite{reimers2019sentence} & 0.043$\blacktriangledown$ & 0.035$\blacktriangledown$ & 0.037$\blacktriangledown$ & 0.061$\blacktriangledown$ & 0.055$\blacktriangledown$ & 0.077$\blacktriangledown$ & 0.257$\blacktriangledown$ & 0.195$\blacktriangledown$ & 0.269$\blacktriangledown$ \\
stsb-distilroberta-base \cite{reimers2019sentence} & 0.045$\blacktriangledown$ & 0.037$\blacktriangledown$ & 0.038$\blacktriangledown$ & 0.061$\blacktriangledown$ & 0.057$\blacktriangledown$ & 0.076$\blacktriangledown$ & 0.260$\blacktriangledown$ & 0.195$\blacktriangledown$ & 0.276$\blacktriangledown$ \\
\midrule
\multicolumn{10}{l}{\textit{\textbf{Bi-Encoder Models}}} \\
ColBERT v2 \cite{khattab2020colbert} (Fine-Tuned) & 0.292 & 0.473$\blacktriangle$ & 0.410$\blacktriangle$ & 0.267$\blacktriangledown$ & 0.455$\blacktriangle$ & 0.451$\blacktriangledown$ & 0.375$\blacktriangle$ & 0.446$\blacktriangle$ & 0.456$\blacktriangle$ \\
DPR \cite{karpukhin-etal-2020-dense} (Zero-Shot) & 0.170$\blacktriangledown$ & 0.300$\blacktriangledown$ & 0.259$\blacktriangledown$ & 0.210$\blacktriangledown$ & 0.309$\blacktriangledown$ & 0.322$\blacktriangledown$ & 0.338$\blacktriangle$ & 0.364 & 0.401$\blacktriangle$ \\
\midrule
\multicolumn{10}{l}{\textit{\textbf{SentenceBERT Bi-Encoders (Zero-Shot)}}} \\
all-MiniLM-L6-v2 \cite{reimers2019sentence} & 0.222$\blacktriangledown$ & 0.394 & 0.334 & \multicolumn{1}{c}{0.264} & \multicolumn{1}{c}{0.419} & \multicolumn{1}{c}{0.419} & \multicolumn{1}{c}{0.365} & \multicolumn{1}{c}{0.412} & \multicolumn{1}{c}{0.420} \\
multi-qa-MiniLM-L6-cos-v1 \cite{reimers2019sentence} & 0.223$\blacktriangledown$ & 0.396 & 0.333 & \multicolumn{1}{c}{0.257} & \multicolumn{1}{c}{0.405} & \multicolumn{1}{c}{0.409} & \multicolumn{1}{c}{0.369} & \multicolumn{1}{c}{0.433} & \multicolumn{1}{c}{0.446} \\
paraphrase-albert-small-v2 \cite{reimers2019sentence} & 0.177$\blacktriangledown$ & 0.341$\blacktriangledown$ & 0.287$\blacktriangledown$ & 0.203$\blacktriangledown$ & 0.316$\blacktriangledown$ & 0.329$\blacktriangledown$ & 0.352$\blacktriangledown$ & \multicolumn{1}{c}{0.392} & 0.410$\blacktriangledown$ \\
msmarco-bert-base-dot-v5 \cite{reimers2019sentence} & 0.234$\blacktriangledown$ & 0.405 & 0.347 & 0.257$\blacktriangledown$ & \multicolumn{1}{c}{0.389} & \multicolumn{1}{c}{0.402} & \multicolumn{1}{c}{0.365} & \multicolumn{1}{c}{0.414} & \multicolumn{1}{c}{0.413} \\
multi-qa-mpnet-base-dot-v1 \cite{reimers2019sentence} & 0.340 & 0.500 & 0.449 & \multicolumn{1}{c}{0.314} & \multicolumn{1}{c}{0.451} & \multicolumn{1}{c}{0.465} & 
 \multicolumn{1}{c}{0.263} & \multicolumn{1}{c}{0.205} & \multicolumn{1}{c}{0.278} \\
\midrule
\multicolumn{10}{l}{\textit{\textbf{LLM-Based (Zero-Shot)}}} \\
BAAI/bge-reranker-v2-gemma \cite{bge_m3} (Pointwise) & 0.266$\blacktriangledown$ & 0.504$\blacktriangle$ & 0.431$\blacktriangle$ & \multicolumn{1}{c}{0.322} & \multicolumn{1}{c}{0.479} & \multicolumn{1}{c}{0.479} & 0.263$\blacktriangledown$ & 0.205$\blacktriangledown$ & 0.278$\blacktriangledown$ \\
INSTRUCTOR \cite{su-etal-2023-one} (Pointwise) & 0.266 & 0.459 & 0.393 & \multicolumn{1}{c}{0.301} & \multicolumn{1}{c}{0.482} & \multicolumn{1}{c}{0.481} & \multicolumn{1}{c}{0.396} & \multicolumn{1}{c}{0.468} & \multicolumn{1}{c}{0.478} \\
RankVicuna \cite{pradeep2023rankvicuna} (Listwise) & 0.296 & 0.467 & 0.400 & \multicolumn{1}{c}{0.315} & \multicolumn{1}{c}{0.446} & \multicolumn{1}{c}{0.459} & 0.382 & \multicolumn{1}{c}{0.426} & \multicolumn{1}{c}{0.456} \\
RankZypher \cite{pradeep2023rankzephyr} (Listwise) & 0.318$\blacktriangle$ & 0.505$\blacktriangle$ & 0.433$\blacktriangle$ & \multicolumn{1}{c}{0.298} & \multicolumn{1}{c}{0.413} & \multicolumn{1}{c}{0.426} & 0.400$\blacktriangle$ & 0.457$\blacktriangle$ & \multicolumn{1}{c}{0.465} \\
\midrule
\multicolumn{10}{l}{\textit{\textbf{Hybrid Models (Trained)}}} \\
BM25InjectionReranker \cite{askari2023injecting} & 0.378$\blacktriangle$ & 0.559$\blacktriangle$ & 0.501$\blacktriangle$ & \multicolumn{1}{c}{0.334} & \multicolumn{1}{c}{0.481} & \multicolumn{1}{c}{0.476} & 0.362$\blacktriangle$ & \multicolumn{1}{c}{0.378} & 0.438$\blacktriangle$ \\
BM25DenseInterpolationReranker \cite{wang2021bert} & 0.340$\blacktriangle$ & 0.500$\blacktriangle$ & 0.449$\blacktriangle$ & \multicolumn{1}{c}{0.314} & \multicolumn{1}{c}{0.451} & \multicolumn{1}{c}{0.465} & \multicolumn{1}{c}{0.361} & \multicolumn{1}{c}{0.433} & \multicolumn{1}{c}{0.376} \\
Coordinate Ascent (MonoBERT + BM25)  & 0.341$\blacktriangle$ & 0.514$\blacktriangle$ & 0.449$\blacktriangle$ & 0.358$\blacktriangle$ & 0.493$\blacktriangle$ & 0.495$\blacktriangle$ & 0.393$\blacktriangle$ & 0.453$\blacktriangle$ & 0.471$\blacktriangle$ \\
Coordinate Ascent (ELECTRA + BM25)  & 0.324$\blacktriangle$ & 0.489$\blacktriangle$ & 0.430$\blacktriangle$ & 0.337$\blacktriangle$ & \multicolumn{1}{c}{0.471} & \multicolumn{1}{c}{0.475} & \multicolumn{1}{c}{0.362} & \multicolumn{1}{c}{0.382} & \multicolumn{1}{c}{0.436} \\
Coordinate Ascent (all-MiniLM-L6-v2 + BM25) & 0.313$\blacktriangle$ & 0.474$\blacktriangle$ & 0.415$\blacktriangle$ & 0.345$\blacktriangle$ & \multicolumn{1}{c}{0.488} & \multicolumn{1}{c}{0.493} & 
 0.385$\blacktriangle$ & 0.445$\blacktriangle$ & 0.460$\blacktriangle$ \\
Coordinate Ascent (ms-marco-MiniLM-L6-v2  + BM25) & 0.323$\blacktriangle$ & 0.480$\blacktriangle$ & 0.424$\blacktriangle$ & 0.349$\blacktriangle$ & \multicolumn{1}{c}{0.489} & \multicolumn{1}{c}{0.494} & \multicolumn{1}{c}{0.376} & 0.421$\blacktriangle$ & 0.439$\blacktriangle$ \\
Coordinate Ascent (BAAI/bge-reranker-v2-gemma  + BM25)  & 0.326$\blacktriangle$ & 0.516$\blacktriangle$ & 0.443$\blacktriangle$ & 0.378$\blacktriangle$ & 0.525$\blacktriangle$ & 0.523$\blacktriangle$ & 0.413$\blacktriangle$ & 0.474$\blacktriangle$ & 0.492$\blacktriangle$ \\
\midrule
TREC Best    
& 0.332$\blacktriangle$           & \multicolumn{1}{c}{--}           & \multicolumn{1}{c}{--} & 0.432$\blacktriangle$	& \multicolumn{1}{c}{--} & 0.612$\blacktriangle$ & \multicolumn{1}{c}{--} & \multicolumn{1}{c}{--} & \multicolumn{1}{c}{--} \\
\midrule
\textbf{REGENT} & \textbf{0.609}$\blacktriangle$ & 
 \textbf{0.785}$\blacktriangle$ & \textbf{0.765}$\blacktriangle$ & \textbf{0.516}$\blacktriangle$ & \textbf{0.627}$\blacktriangle$ & \textbf{0.645}$\blacktriangle$ & \textbf{0.498}$\blacktriangle$ & 0.423$\blacktriangle$ & \textbf{0.559}$\blacktriangle$ \\
\bottomrule
\end{tabular}
}
\end{table*}

\begin{table}[t]
\centering
\caption{Results on \textbf{TREC Robust 2004 title queries} comparing REGENT to full retrieval methods. Note: REGENT is \textit{reranking} the BM25 full retrieval results presented at the top. 
%$\blacktriangle$ denotes significant improvement and $\blacktriangledown$ denotes significant deterioration compared to BM25+RM3 (denoted $\star$) using a paired t-test at $p<0.05$.
}
\label{tab:res-tab-full-ret}
\scalebox{0.87}{
% Using 'siunitx' S column to align numbers by decimal point, inspired by the provided example.
% table-format=1.3 aligns numbers with 1 digit before and up to 3 after the decimal.
% table-space-text-post reserves space for the triangle symbol, ensuring consistent column width.
\begin{tabular}{
  @{}l
  S[table-format=1.2, table-space-text-post={$\blacktriangle$}]
  S[table-format=1.2, table-space-text-post={$\blacktriangle$}]
  S[table-format=1.2, table-space-text-post={$\blacktriangle$}]
  @{}
}
\toprule
% Headers for S-type columns must be wrapped in braces {} to be treated as text.
\textbf{Method} & {\textbf{MAP}} & {\textbf{nDCG@20}} & {\textbf{P@20}} \\
\midrule
BM25 & 0.29 & 0.44 & 0.38 \\
\midrule
BERT-QE  \cite{zheng-etal-2020-bert} & 0.39$\blacktriangle$ & 0.55$\blacktriangle$ & 0.49$\blacktriangle$ \\
CEQE \cite{naseri2021ceqe} & 0.31$\blacktriangle$ & 0.46$\blacktriangle$ & 0.40$\blacktriangle$ \\
DeepCT \cite{dai2020context} & 0.20$\blacktriangledown$ & 0.38$\blacktriangledown$ & 0.33$\blacktriangledown$ \\
NPRF \cite{li-etal-2018-nprf} \cite{li-etal-2018-nprf} & 0.29 & 0.45 & 0.41 \\
EQFE \cite{dalton2014entity} & 0.33 & 0.43 & 0.38 \\
ANCE-MaxP \cite{xiong2020ance} & 0.13$\blacktriangledown$ & 0.31$\blacktriangledown$ & 0.25$\blacktriangledown$ \\
BERT-MaxP \cite{dai2019deeper} & 0.31 & 0.48 & 0.42 \\
TCT-ColBERT \cite{lin2020distilling} & 0.23 & 0.43 & 0.36 \\
PARADE \cite{li2020parade} & 0.30 & 0.52 & 0.45 \\
SPLADE v2 \cite{thibault2021splade} & 0.22$\blacktriangledown$ & 0.42 & 0.36 \\
\midrule
% Use \bfseries for bolding numbers within an S column.
REGENT & \textbf{0.61}$\blacktriangle$ & \textbf{0.79}$\blacktriangle$ & \textbf{0.77}$\blacktriangle$ \\
\bottomrule
\end{tabular}
}
\end{table}

%\input{tex-files/tables/ablations}

% \subsection{Overall Results}
% \label{subsec:Overall Results}
% \begin{table}[t]
%     \centering
%     \caption{Ablation study for architectural choices. Results reported on TREC Core 2018.}
%     \scalebox{0.8}{
%     \begin{tabular}{lccc}
%         \toprule
%         \textbf{Model Variant} & \textbf{MAP} & \textbf{nDCG@20} & \textbf{P@20} \\
%         \midrule
%         No Entities & 0.134 & 0.174 & 0.189 \\
%         No BM25 & 0.449 & 0.561 & 0.605 \\
%         \bottomrule
%     \end{tabular}
%     }
%     \label{tab:architecture-ablation-study}
% \end{table}

\begin{table}[t]
\centering
\caption{Ablation study for architectural choices. Results reported on TREC Core 2018.}    \label{tab:architecture-ablation-study}

%\scalebox{0.8}{
\begin{tabular}{
  @{}l
  S[table-format=1.3, table-space-text-post={$\blacktriangle$}]
  S[table-format=1.3, table-space-text-post={$\blacktriangle$}]
  S[table-format=1.3, table-space-text-post={$\blacktriangle$}]
  @{}
}
\toprule
\textbf{Model Variant} & \textbf{MAP} & \textbf{nDCG@20} & \textbf{P@20} \\
\midrule
RENET(No Entities)         & 0.134$\blacktriangledown$ & 0.174$\blacktriangledown$ & 0.189$\blacktriangledown$ \\
REGENT(No BM25)  & 0.449$\blacktriangledown$ & 0.561$\blacktriangledown$ & 0.605$\blacktriangledown$ \\
\midrule
\textbf{REGENT(Full)} & \textbf{0.516}& \textbf{0.627} & \textbf{0.645}  \\
\bottomrule
\end{tabular}
%}
\end{table}

As shown in Table \ref{tab:results}, \textsc{REGENT} decisively establishes a new state-of-the-art for re-ranking on all three benchmarks. Performance gains are most pronounced on TREC Robust04, where \textsc{REGENT} achieves a MAP of 0.609, a remarkable 108\% relative improvement over the strong baseline BM25, and an nDCG@20 of 0.785. 

Crucially, these gains are not limited to a single class of models. \textsc{REGENT}T consistently surpasses a wide array of strong baselines, including fine-tuned cross-encoders like DeBERTa, multi-vector models like ColBERT v2, recent zero-shot LLM re-rankers such as RankZephyr, and other state-of-the-art hybrid models. This consistent and significant outperformance provides strong evidence that \textsc{REGENT}'s fine-grained integration of lexical signals and entity context offers a more effective path to relevance modeling than what is possible with existing architectures.

To provide broader context for our results on the historically significant TREC Robust 2004 benchmark, we also compare \textsc{REGENT}'s re-ranking performance against several influential full retrieval (first-stage) models in Table\ref{tab:res-tab-full-ret}. While re-rankers and full retrievers serve different functions in a multi-stage system, this comparison highlights the substantial effectiveness gains \textsc{REGENT} provides over the initial BM25 ranking it refines. It demonstrates that our re-ranking approach elevates performance far beyond what classic, standalone retrieval methods could achieve on this challenging task.

\textbf{\textit{In the following sections, we analyze and discuss the results for the TREC Core 2018 dataset due to lack of space but similar results were obtained on all datasets.}}

\subsection{Effect of Model Components}
\label{subsec:Impact of Model Components}

We first ask: \textbf{RQ1: What is the individual contribution and synergistic effect of the entity-based semantic pathway and token-level lexical guidance within \textsc{REGENT}'s relevance-guided attention mechanism?} To answer this, we conducted a series of ablation studies. The results (Table~\ref{tab:architecture-ablation-study}) confirm that both the entity-based semantic pathway and the token-level lexical guidance are critical, but they play distinct and complementary roles in the model's success. 

\textbf{The Primacy of the Semantic Skeleton.} First, we assessed the individual contributions of the entity and BM25 signals. Removing the entity information pathway entirely caused a catastrophic performance drop, with MAP decreasing from 0.516 to 0.134 (–74\%).\footnote{While one might expect this variant to perform similarly to a standard cross-encoder, its architecture is not identical. It retains the cross-attention and fusion mechanism structures of \textsc{REGENT}, which are not optimized to function without the entity pathway, leading to sub-optimal performance compared to a purpose-built cross-encoder like MonoBERT.} This is by far the most significant impact of any single component, validating our central hypothesis that the ``semantic skeleton'' provided by the entity context is not merely an auxiliary feature but the primary driver of \textsc{REGENT}’s ability to perform deep semantic reasoning. Without it, the model loses its ability to understand the conceptual landscape of a document.

In contrast, removing only the token-level BM25 signal also led to a notable performance decline (MAP from 0.516 to 0.449). Although smaller, this drop indicates that fine-grained lexical signals provide valuable complementary information that the entity pathway alone cannot capture. This suggests that the BM25 signal acts as a powerful ``fine‑tuner,'' helping the model ground its high-level semantic understanding in specific evidence-bearing term matches.

\textbf{Granularity is Crucial: Token‑Level vs. Document‑Level BM25.} To further validate our architectural choice of fine‑grained signal integration, we investigated the impact of using a single document‑level BM25 score instead of our token‑level approach. In this variant, the model first computes a neural score using the entity‑aware attention layers (with token‑level BM25 disabled), then combines this score with the document’s overall BM25 score using a learned linear transformation: $\text{score} = W [s_{\text{neural}}; s_{\text{BM25}}]$.

This experiment allows us to directly assess how the granularity of the BM25 signal affects performance. The results are striking: the document‑level approach leads to a substantially lower MAP of 0.418. More revealingly, this is even worse than the model with no BM25 signal at all (MAP 0.449). This finding suggests that a single coarse-grained BM25 score, when naively aggregated, can be counterproductive, potentially confusing the model more than helping it. It provides strong evidence that for a lexical signal to be effective in a multi‑vector attention framework, it must be integrated at the token level where it can directly guide fine-grained interactions.

\textbf{Takeaway.} Taken together, these ablations paint a clear picture. The entity pathway provides the foundational semantic understanding, while token-level BM25 integration provides precise lexical guidance. While both are important, the model's strength lies in processing them at the right granularity and fusing them effectively. Crucially,  the sharp performance gap---74\% drop without entities vs. 13\% without BM25---shows that entity signals form a robust semantic backbone, retaining most of the model's performance on their own. This highlights that semantic understanding is essential for complex retrieval, and lexical matching alone is insufficient.
%Crucially, the stark asymmetry in performance drops (–74\% without entities vs –13\% without BM25) reveals that the entity-based ``semantic skeleton'' alone can provide a complete conceptual picture for relevance decisions, retaining 87\% of the full model's performance. This demonstrates that entities successfully capture the document's core semantic structure and can function as a self-sufficient relevance framework. In contrast, lexical signals without semantic context prove fundamentally inadequate, suggesting that sophisticated term matching cannot substitute for high-level conceptual understanding in complex retrieval scenarios.

\subsection{Effect of Fusion Mechanism}
\label{subsec:Abalation Study: Effect of Fusion Mechanism}

\begin{table}[t]
\centering
\caption{Ablation study for fusion choices. Results reported on TREC Core 2018.}
\label{tab:fusion_ablation}
%\scalebox{0.8}{
\begin{tabular}{
  @{}l
  S[table-format=1.3, table-space-text-post={$\blacktriangle$}]
  S[table-format=1.3, table-space-text-post={$\blacktriangle$}]
  S[table-format=1.3, table-space-text-post={$\blacktriangle$}]
  @{}
}
\toprule
\textbf{Method} & \textbf{MAP} & \textbf{nDCG@20} & \textbf{P@20} \\
\midrule
REGENT (Additive)         & 0.497$\blacktriangledown$ & 0.619 & 0.646 \\
REGENT (Attention-based)  & 0.384$\blacktriangledown$ & 0.491 & 0.494 \\
REGENT (Equal weighting)  & 0.496$\blacktriangledown$ & 0.618 & 0.644 \\
REGENT (Gated GELU)       & 0.502 & 0.628 & 0.648 \\
REGENT (Hard switch)      & 0.426$\blacktriangledown$ & 0.565$\blacktriangledown$ & 0.600 \\
REGENT (Learned tanh)     & 0.498 & 0.617 & 0.655 \\
\midrule
\textbf{REGENT (Learned Sigmoid)} & \textbf{0.516}& \textbf{0.627} & \textbf{0.645}  \\
\bottomrule
\end{tabular}
%}
\end{table}

\textbf{RQ2: How does the choice of fusion mechanism impact \textsc{REGENT}'s performance, and to what extent does its effectiveness depend on a specific fusion strategy versus its dual-pathway architecture?} To answer this, we replaced our default learned sigmoid gate with six alternative fusion strategies and re-trained our model: (1) \textbf{Gated GELU}, employing a sophisticated nonlinear gating network with GELU activation and dropout regularization; (2) \textbf{Additive}, using straightforward element-wise addition followed by normalization; (3) \textbf{Equal Weighting}, applying a fixed 50-50 average of token and entity outputs; (4) \textbf{Learned Tanh}, implementing tanh-based gating as an alternative to sigmoid activation; (5) \textbf{Hard Switch}, utilizing a binary selector that favors entity attention when entities are present; and (6) \textbf{Attention-based}, leveraging multi-head attention over stacked token and entity representations.

The results reveal a remarkable consistency across these diverse methods, with overall MAP scores clustering within a tight $\approx$1\% margin (0.496 to 0.502, see Table~\ref{tab:fusion_ablation}). This narrow variance provides powerful evidence that \textsc{REGENT}'s effectiveness stems primarily from its dual-pathway architecture itself---which successfully separates lexical and semantic processing---rather than from a specific, highly-tuned fusion technique.

However, a deeper analysis reveals that this macro-level stability masks important specializations. No single fusion method was universally optimal. Instead, different query types and difficulty levels responded best to different strategies, pointing toward the potential of adaptive routing. Our analysis shows that a query-specific fusion selection could yield substantial improvements---up to 41.5\% for medium-difficulty queries. The robustness of this difficulty-based analysis is supported by strong correlations between WIG scores and performance (0.67--0.71), validating our classification approach.

\textbf{Takeaway.} The key finding is that \textsc{REGENT}'s performance is fundamentally robust to the choice of fusion mechanism, confirming that the dual-pathway design is the core contribution. Furthermore, the fact that different query types favor different fusion strategies suggests a promising and quantifiable avenue for future work in query-adaptive routing, where the fusion method could be selected dynamically based on query characteristics.

\subsection{Effect of Supervised Entity Ranker}
\label{subsec:Effect of Supervised Entity Ranker}

\textbf{RQ3: Is the supervised BERT-based entity ranker (Section \ref{subsec:Query-Specific Entity Set Construction}) a necessary component for \textsc{REGENT}'s performance, or can simpler (unsupervised or less complex supervised) entity scoring methods achieve comparable results?} To answer this, we replaced our main entity ranker with four alternatives: (1) \textbf{BM25}: An unsupervised lexical approach where candidate entities are scored by running the query against a corpus of their textual descriptions (from DBpedia), (2) \textbf{MaxSim}: An unsupervised semantic approach where candidate entities are scored by their maximum (Wikipedia2Vec) cosine similarity to the entities linked to the query, (3) \textbf{CentroidSim}: An unsupervised method that scores each entity by its cosine similarity to the average query entity embedding (centroid), and (4) \textbf{LogReg}: A simpler supervised approach using a logistic regression classifier instead of a full BERT-based ranker. We then used the entity sets obtained using each of these methods to train and evaluate our full \textsc{REGENT} model. 

\begin{table}[t]
\centering
\caption{Ablation study for entity scoring choices. Results reported on TREC Core 2018.}
\label{tab:entity_ranking_ablation}
\scalebox{0.8}{
\begin{tabular}{
  @{}l
  S[table-format=1.3, table-space-text-post={$\blacktriangle$}]
  S[table-format=1.3, table-space-text-post={$\blacktriangle$}]
  S[table-format=1.3, table-space-text-post={$\blacktriangle$}]
  @{}
}
\toprule
\textbf{Method} & \textbf{MAP} & \textbf{nDCG@20} & \textbf{P@20} \\
\midrule
RENET(BM25)         & 0.132$\blacktriangledown$ & 0.179$\blacktriangledown$ & 0.207$\blacktriangledown$ \\
REGENT(MaxSim)  & 0.071$\blacktriangledown$ & 0.121$\blacktriangledown$ & 0.145$\blacktriangledown$ \\
REGENT(CentroidSim)  & 0.081$\blacktriangledown$ & 0.121$\blacktriangledown$ & 0.149$\blacktriangledown$ \\
REGENT(LogReg)  & 0.093$\blacktriangledown$ & 0.222$\blacktriangledown$ & 0.213$\blacktriangledown$ \\
%CentroidSim       & 0.502 & 0.628 & 0.648 \\
%QLD      & 0.426$\blacktriangledown$ & 0.565$\blacktriangledown$ & 0.600 \\
\midrule
\textbf{REGENT(BERT)} & \textbf{0.516}& \textbf{0.627} & \textbf{0.645}  \\
\bottomrule
\end{tabular}
}
\end{table}
The results, presented in Table \ref{tab:entity_ranking_ablation}, show that replacing our supervised entity ranker with any simpler method leads to a complete collapse in performance. The MAP score plummets from 0.516 to as low as 0.093---a drop of nearly 82\%. Neither unsupervised methods (BM25, MaxSim, CentroidSim) nor a simpler supervised model (LogReg) could produce an entity signal useful for \textsc{REGENT}'s downstream reasoning. This finding underscores that the model's success is not from using any entity information, but is critically dependent on the high-quality, contextually-aware relevance scores that only the sophisticated, BERT-based ranker can provide. Because these high-quality scores are intrinsically linked to the scaling operation, this result also justifies our focus on evaluating the entire entity pipeline rather than attempting to isolate the scaling effect alone.

\textbf{Takeaway.} The complexity of our supervised entity selection pipeline is not just justified; it is essential. The ability to accurately identify and score the most relevant entities for a given query is a prerequisite for \textsc{REGENT}'s success and a core component of its state-of-the-art performance.

\subsection{Feature Pipeline vs. Model Design}
\label{subsec:Feature Pipeline vs. Model Design}

\textbf{RQ4: What is the primary driver of \textsc{REGENT}'s effectiveness: the supervised entity selection pipeline or its novel architectural design for processing those features?} To answer this, we designed a controlled experiment to isolate each contribution. We created \textbf{BERT-Entity}, a stronger baseline that receives exactly the same relevance-scored entity features as \textsc{REGENT}, but processes them through a standard cross-encoder architecture via simple feature concatenation. This creates a clear three-way comparison.

\textbf{The Impact of the Entity Pipeline.} First, we measured the effect of our entity selection pipeline on its own. Simply providing these high-quality entity features to a standard BERT model yields substantial improvements, boosting MAP from 0.302 to 0.390 on TREC Core 2018---a 29.1\% relative gain. This result confirms that our supervised entity selection pipeline is a highly effective method for representing documents for neural ranking.

\textbf{The Added Value of the \textsc{REGENT} Architecture.} Next, we compared \textsc{REGENT} to the much stronger BERT-Entity baseline. Since both models receive identical inputs, this comparison isolates the architectural advantage. \textsc{REGENT} achieves another significant 32.3\% relative improvement in MAP (from 0.390 to 0.516). This demonstrates that \emph{how} signals are processed is as crucial as \emph{what} signals are available. \textsc{REGENT}'s relevance-guided attention, which deeply integrates entity and token reasoning, proves architecturally superior to the simpler approach of feature concatenation.

\textbf{Takeaway.} \textsc{REGENT}'s state-of-the-art performance emerges from the synergy of two distinct and significant contributions: an effective, supervised approach to entity-aware document representation, and a novel architecture designed to reason with these representations. The results show that while the entity pipeline provides a powerful foundation, the full potential is only realized through \textsc{REGENT}'s specialized attention mechanism.

\subsection{Analysis by Query Characteristics}
\label{subsec:Analysis by Query Characteristics}

% \begin{figure*}[t]
% \centering
% \includegraphics[width=\textwidth]{images/wig_analysis.png}
% \caption{WIG-based performance analysis showing REGENT's effectiveness on difficult queries. 
% (Top-left) MAP performance by query difficulty shows transformative improvements over SBERT cross-encoder. 
% (Top-right) Consistent advantages across MAP and MRR metrics. 
% (Bottom-left) WIG score validation: lower WIG correlates with higher REGENT advantage. 
% (Bottom-right) Query distribution ensures balanced evaluation.}
% \label{fig:wig_analysis}
% \end{figure*}

\begin{figure*}[tb]
\centering
\begin{subfigure}{0.48\textwidth}
   \centering
   \includegraphics[width=\textwidth]{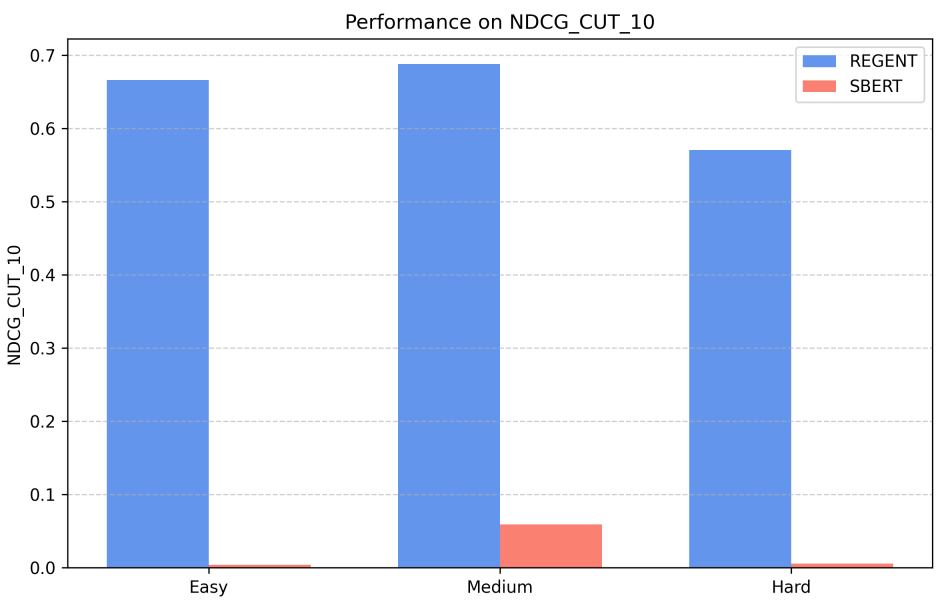}
   \caption{Performance comparison of REGENT and SBERT, segmented by query difficulty determined via our WIG-based classifier. SBERT's effectiveness diminishes on harder queries while REGENT's remains high. Hence, REGENT's computational overhead is most impactful on the most challenging queries.}
   \label{fig:wig_analysis}
\end{subfigure}
\hfill
\begin{subfigure}{0.48\textwidth}
   \centering
   \includegraphics[width=\textwidth]{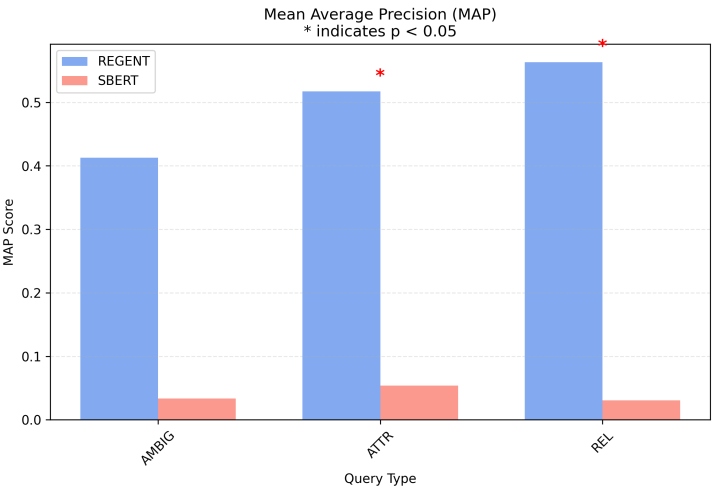}
   \caption{MAP of REGENT and a Sentence-BERT baseline, segmented by the query's semantic type. As hypothesized, REGENT's performance advantage grows with the relational complexity of the query, from Ambiguous/Keyword (AMBIG) to Attributive (ATTR) and is most pronounced on Relational (REL) queries that involve interactions between multiple entities.}
   \label{fig:query_type_analysis}
\end{subfigure}
\caption{Query-level analysis on TREC Core 2018.}
\label{fig:query_type_analysis}
\end{figure*}

% \begin{figure}[t]
%   \centering
%   \includegraphics[width=\linewidth]{images/wig_analysis.png} % Replace with the actual path to your figure
%   \caption{Performance comparison of REGENT and SBERT on Core18, segmented by query difficulty determined via our WIG-based classifier. nDCG@10 scores, where SBERT's effectiveness diminishes on harder queries while REGENT's remains high. This validates that REGENT's computational overhead is most impactful on the most challenging queries.}
%   \label{fig:wig_analysis}
% \end{figure}

\textbf{RQ5: How does \textsc{REGENT}'s performance vary across different types of query semantic complexity?} To answer this, we analyzed results across three query categories: (1) Relational (REL)--queries seeking relationships between entities; (2) Attributive (ATTR)--queries about properties of a single entity; and (3) Ambiguous/Keyword (AMBIG)--broad or definitional queries.
We adopted a two-stage annotation process for classifying all 50 topics. First, two state-of-the-art LLMs (Gemini 2.5 Pro and Claude Sonnet 4) independently labeled the queries. Their agreement yielded a Cohen’s Kappa of 0.42 (``Moderate Agreement''); 11 of the 12 disagreements occurred between ATTR and REL, indicating a nuanced semantic overlap. In the second stage, the authors adjudicated all labels to establish a gold-standard annotation.

The results provide a clear validation of our hypothesis. \textsc{REGENT}'s performance advantage grows precisely as the query's semantic complexity increases. \textsc{REGENT}'s gains over SBERT, chosen as a representative and strong sentence-transformer baseline, are largest for REL queries (+0.533 MAP, $n=10$, $p{<}0.05$), followed by ATTR (+0.463 MAP, $n=35$, $p{<}0.05$), and AMBIG (+0.379 MAP, $n=5$, $p{<}0.05$). Improvements are particularly strong for queries straddling the ATTR–REL boundary, where \textsc{REGENT}'s joint entity-relationship modeling proves most beneficial.

\textbf{Takeaway.} \textsc{REGENT} excels at modeling entity-centric queries, especially those involving complex relationships. Its performance advantage scales with the semantic richness of the query—confirming that relevance-aware entity modeling is central to its effectiveness.

\subsection{Analysis by Query Difficulty}
\label{subsec:Performance Across Query Difficulty}

\begin{figure*}[t]
    \centering
    \includegraphics [width=\textwidth]{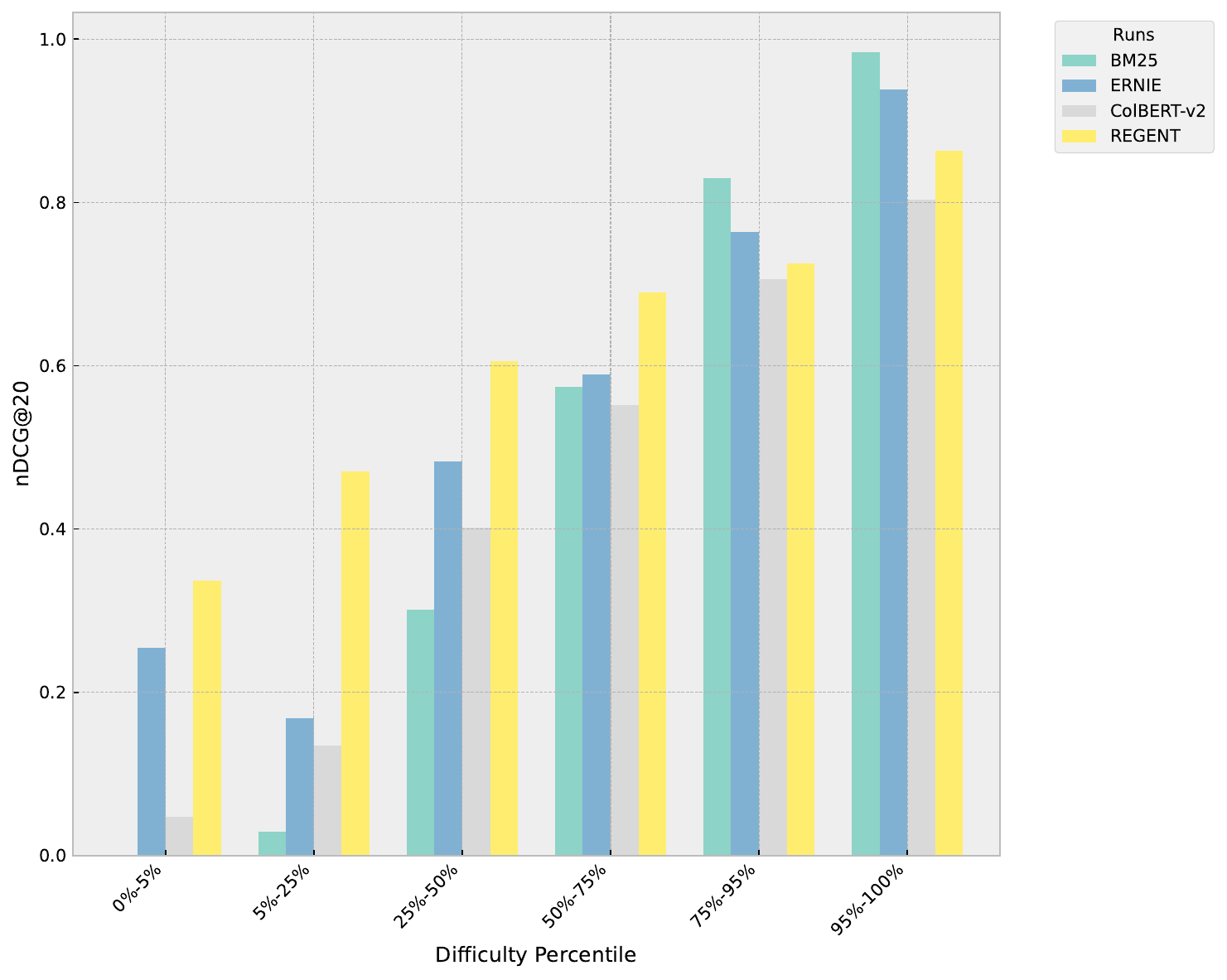}
    \caption{Difficulty test on Core18. 5\% most difficult queries for BM25 to the left and the 5\% easiest ones to the right. Performance reported as macro-averages across queries.
    }
    \label{fig:diff-test}
\end{figure*}

\textbf{RQ6: How does \textsc{REGENT}'s performance vary across different levels of query difficulty, particularly for queries where initial lexical retrieval methods (like BM25) fail?} To answer this\footnote{We use minir-plots for this. See: \url{https://github.com/laura-dietz/minir-plots}}, we first measured each query's difficulty. We define a query's difficulty based on the performance of the initial BM25 ranking (nDCG@20); queries with low initial BM25 scores are considered ``hard'' (Figure~\ref{fig:diff-test}). For the most difficult queries (0-5th percentile), where BM25 completely fails to retrieve relevant documents (nDCG@20 $\approx$ 0), \textsc{REGENT} demonstrates a unique capability. While other strong baselines like ColBERT-v2, selected to represent the multi-vector paradigm,  offer only modest improvements, \textsc{REGENT} ``rescues'' these failed queries, achieving a respectable nDCG@20 of 0.35.
This trend is not an anomaly. Across all challenging query bins (0-75th percentile), \textsc{REGENT} consistently establishes a substantial performance gap over all other methods. For instance, in the 5-25th percentile, it more than doubles the nDCG@20 of its closest competitor. As expected, for the easiest queries (95-100th percentile) where lexical signals are already strong, all models perform well. However, it is \textsc{REGENT}'s exceptional performance on the most challenging queries that validates its architectural design.

\textbf{Takeaway.} \textsc{REGENT}'s advantage is most pronounced when simple lexical matching is insufficient. Its ability to turn failed queries into successful ones provides strong evidence that its relevance-guided attention mechanism, leveraging both entity context and token-level signals, offers a more robust path to understanding relevance than architectures more reliant on lexical overlap.

\subsection{Distribution of Relevant Documents}
\label{subsec:Distribution of Relevant Documents}

\textbf{RQ7: How effectively does \textsc{REGENT} re-structure the initial document ranking by promoting relevant documents, especially highly relevant ones and those missed by lexical models?}
%Beyond aggregate metrics, \textsc{REGENT}'s strength is further validated by its ability to fundamentally restructure the initial ranking. It demonstrates exceptional precision by promoting relevant documents from deep within the candidate set to the top positions. 
Even for queries where BM25 struggles, \textsc{REGENT} places 374 relevant documents within the top 10, and successfully surfaces nearly half of all relevant content (1,252 documents) within the top 50. The impact is even more pronounced for the most critical documents. \textsc{REGENT} dramatically improves the average rank of highly relevant documents (NIST grade 2) from 171 to 113. This powerful promotion effect, consistent across all query types, demonstrates that \textsc{REGENT}'s architecture is not merely refining the top of the list, but actively identifying and elevating high-value documents that simple lexical models miss entirely.

\textbf{Takeaway.} \textsc{REGENT}'s success is not just from improving scores, but from its ability to correct the initial ranking. It consistently finds the ``needles in the haystack''---highly relevant documents buried deep in the initial candidate set---and promotes them to the top, which is a key requirement for effective retrieval on complex information needs.

\subsection{Computational Cost Analysis}
\label{sec:computational_cost}

\textbf{RQ8: What is the computational cost of \textsc{REGENT} compared to baselines, and is this overhead justified by its performance gains, especially for queries of varying difficulty? } Although not our primary focus, we compared \textsc{REGENT}'s runtime to a SentenceBERT cross-encoder on Core18. A query–document pair is processed by \textsc{REGENT} in 25.5\,ms---roughly $3\times$ slower than SBERT (8.5\,ms). To assess whether this overhead is justified, we used Weighted Information Gain (WIG)~\cite{cronen2002predicting}, a pre-retrieval query performance predictor, to classify queries into difficulty bins. Unlike using the post-retrieval BM25 score, this a priori method is useful for designing adaptive systems that could selectively deploy \textsc{REGENT} on queries predicted to be difficult.

Our results show that \textsc{REGENT}'s gains grow sharply with query difficulty. On hard queries, MAP jumps from 0.023 (SBERT) to 0.319 (\textbf{+1,316\%}); for medium queries, from 0.070 to 0.579 (\textbf{+727\%}); and even on easy queries, from 0.050 to 0.654.  These results (Figure~\ref{fig:wig_analysis}) highlight \textsc{REGENT}'s value precisely where traditional models fail.

\textbf{Takeaway.} \textsc{REGENT}'s computational cost is a strategic investment. For efficiency-conscious applications, a hybrid approach could deploy \textsc{REGENT} selectively on difficult queries where its order-of-magnitude improvements are most needed. The overhead is justified not despite its cost, but because it enables solving retrieval problems that other methods cannot handle effectively.

\subsection{Comparison to Other Entity-Aware Baselines}
\label{subsec:Comparison to Other Entity-Aware Baselines}

A key result is \textsc{REGENT}'s large performance margin over other entity-aware baselines like EDRM. We attribute this to our \emph{dynamic, query-specific entity processing}. EDRM relies on a static, pre-computed entity similarity matrix, treating entity relationships as context-agnostic. In contrast, \textsc{REGENT}'s supervised entity ranker and attention mechanism learn to identify and weight entities based on the specific query's context. This ability to dynamically reason about entity relevance appears crucial for handling the complex information needs present in our evaluation datasets

\subsection{Case Study Analysis: Sony Cyberattack} 
\label{subsec:Case Study Analysis}

% To illustrate \textbf{\textsc{REGENT}}'s effectiveness, we analyze two queries where it dramatically outperforms lexical baselines by understanding context and relationships. First, for the querry ``\textit{Wildlife Extinction}''(Query ID: 347), which  explicitly asks for conservation efforts in countries \emph{other than the U.S.}, \textsc{REGENT} accurately captures the geographic constraint. A document discussing U.S. spotted owl conservation receives a perfect relevance score of 1.0 from BM25 due to strong keyword overlap. In contrast, \textsc{REGENT} leverages entity context to detect the document's U.S. focus and assigns a near-zero score of \textbf{0.0014}, showcasing its ability to handle nuanced query narratives.
% %
% Second, for the query ``\textit{Sony Cyberattack}'' (Query ID: 822) which seeks the \emph{specific} group responsible, \textsc{REGENT} assigns a high relevance score of \textbf{0.982} to a document about a White House information leak involving Russia, despite a BM25 score of 0.0 and the absence of the term ``Sony.'' \textsc{REGENT} identifies the conceptual pattern of a state-sponsored information leak using related entities and topics, demonstrating semantic reasoning beyond lexical cues. This illustrates \textsc{REGENT}'s capacity to retrieve documents based on latent relationships, not just surface-level keyword matches.

\begin{figure*}[t]
    \centering
    \includegraphics [width=\textwidth]{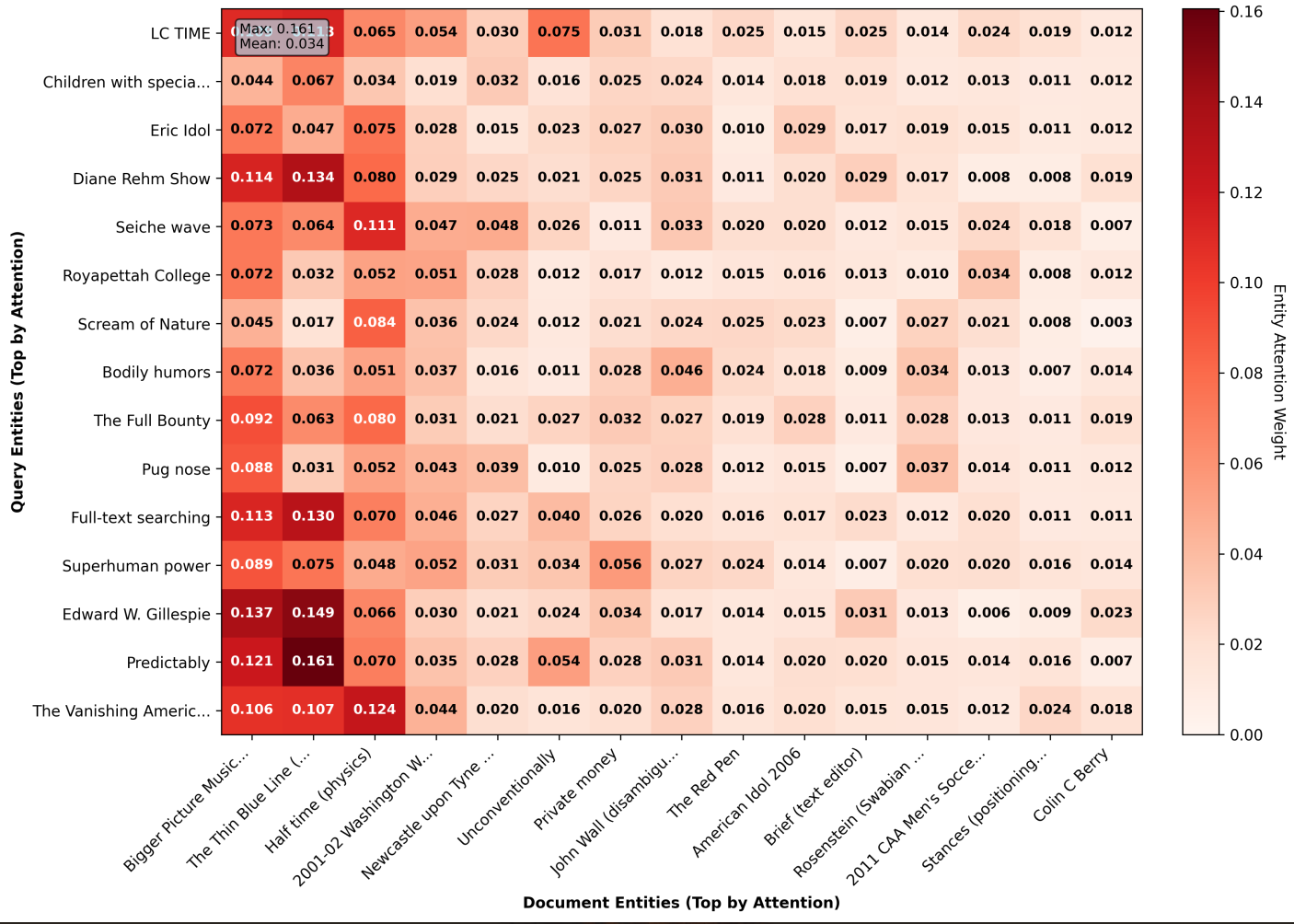}
    \caption{Visualization of entity attention patterns for query ``\textit{Sony Cyberattack}''.
    }
    \label{fig:entity-attention}
\end{figure*}

To demonstrate how \textsc{REGENT} enables deep semantic reasoning, we examine its performance on a challenging query (QueryID: 342, Type: ATTR) where traditional lexical models fail. The query seeks to identify the specific group responsible for the Sony Pictures cyberattack; generic country-level references are considered non-relevant. A document discussing the White House's response to the WannaCry ransomware receives a BM25 score of 0.0, as it lacks the keyword ``cyberattack.'' \textsc{REGENT}, however, ranks it highly (score: 0.982). Its reasoning begins with entity-to-entity attention, where it finds no direct match for ``Sony'' but identifies a strong thematic link between query entities like ``\texttt{Full-text searching}'' (a forensic technique) and document entities like ``\texttt{The Thin Blue Line}'' (associated with investigations). As shown in Figure~\ref{fig:entity-attention}, rather than indicating a direct relation, the high attention score reflects a shared thematic context of ``Investigation and Attribution,'' aligning the query's goal ---- attributing responsibility---with the document's content on leaks, hacks, and attribution. The context captured by $A_e$ informs the entity-to-token attention module ($A_{et}$), which highlights key evidence---e.g., linking the ``crippling hack on Sony Pictures in 2014'' to ``Lazarus'' despite no exact lexical match. The adaptive fusion mechanism then prioritizes the entity pathway, as weak lexical signals are outweighed by strong semantic cues.

\textbf{Takeaway.}  This case illustrates how \textsc{REGENT} surfaces deeply relevant documents that traditional models miss. By constructing a semantic context through entity relationships, it moves beyond simple term matching and captures nuanced aspects of user intent, even when lexical overlap is minimal.

\section{Conclusion}
\label{sec:Conclusion}

We present \textsc{REGENT}, a multi-vector neural re-ranking model that advances integration of traditional IR signals and entity knowledge within neural architectures. At its core is \emph{relevance-guided attention}, a novel mechanism that dynamically fuses lexical and semantic cues to guide attention computations, moving beyond uniform token interactions. Across three large-scale datasets featuring long, information-rich documents and complex queries, \textsc{REGENT} delivers state-of-the-art performance, outperforming BM25 by up to 108\% and surpassing strong baselines like RankT5, ColBERT, and even LLM-based re-rankers such as RankVicuna. Critically, removing the entity-aware pathway leads to a catastrophic 74\% drop in performance, validating our hypothesis that entities serve as a vital ``semantic skeleton'' for understanding document relevance.

Our work makes three fundamental contributions to neural IR: (1) We demonstrate that token-level BM25 integration significantly outperforms document-level approaches, providing strong evidence that lexical signals must be integrated at the appropriate granularity to be effective in multi-vector attention frameworks. (2) We show that entity information—traditionally confined to single-vector models—can be effectively integrated into multi-vector architectures through our dual-pathway attention design, opening new avenues for entity-aware neural retrieval. (3) We introduce the first ranking-guided attention mechanism that dynamically fuses lexical matching with entity-level semantics, moving beyond static signal combination toward contextual relevance modeling.

\textsc{REGENT} marks a meaningful step toward retrieval systems capable of human-like semantic reasoning. By unifying traditional IR strengths with modern neural architectures, it delivers both strong empirical gains and a solid foundation for future advances in entity-aware neural IR. While we focused on a specific instantiation of relevance-guided attention, future work could explore alternative integration methods (e.g., gating) or attention variants to further build on this paradigm. Our work focuses on English-language retrieval, and adapting the token-level BM25 integration for languages without clear word boundaries, such as CJK, remains an important area for future work.

\bibliographystyle{ACM-Reference-Format}
%\balance
\bibliography{references}


\begin{thebibliography}{74}

%%% the Web address.

\ifx \showCODEN    \undefined \def \showCODEN     #1{\unskip}     \fi
\ifx \showDOI      \undefined \def \showDOI       #1{#1}\fi
\ifx \showISBNx    \undefined \def \showISBNx     #1{\unskip}     \fi
\ifx \showISBNxiii \undefined \def \showISBNxiii  #1{\unskip}     \fi
\ifx \showISSN     \undefined \def \showISSN      #1{\unskip}     \fi
\ifx \showLCCN     \undefined \def \showLCCN      #1{\unskip}     \fi
\ifx \shownote     \undefined \def \shownote      #1{#1}          \fi
\ifx \showarticletitle \undefined \def \showarticletitle #1{#1}   \fi
\ifx \showURL      \undefined \def \showURL       {\relax}        \fi
% invisible to TeX
\providecommand\bibfield[2]{#2}
\providecommand\bibinfo[2]{#2}
\providecommand\natexlab[1]{#1}
\providecommand\showeprint[2][]{arXiv:#2}

\bibitem[Akkalyoncu~Yilmaz et~al\mbox{.}(2019)]%
        {akkalyoncu-yilmaz-etal-2019-cross}
\bibfield{author}{\bibinfo{person}{Zeynep Akkalyoncu~Yilmaz}, \bibinfo{person}{Wei Yang}, \bibinfo{person}{Haotian Zhang}, {and} \bibinfo{person}{Jimmy Lin}.} \bibinfo{year}{2019}\natexlab{}.
\newblock \showarticletitle{Cross-Domain Modeling of Sentence-Level Evidence for Document Retrieval}. In \bibinfo{booktitle}{\emph{Proceedings of the 2019 Conference on Empirical Methods in Natural Language Processing and the 9th International Joint Conference on Natural Language Processing (EMNLP-IJCNLP)}}. \bibinfo{publisher}{Association for Computational Linguistics}, \bibinfo{address}{Hong Kong, China}, \bibinfo{pages}{3490--3496}.
\newblock
\urldef\tempurl%
\url{https://doi.org/10.18653/v1/D19-1352}
\showDOI{\tempurl}


\bibitem[Allan et~al\mbox{.}(2017)]%
        {allan2017trec}
\bibfield{author}{\bibinfo{person}{James Allan}, \bibinfo{person}{Donna Harman}, \bibinfo{person}{Evangelos Kanoulas}, \bibinfo{person}{Dan Li}, \bibinfo{person}{Christophe Van~Gysel}, {and} \bibinfo{person}{Ellen~M Voorhees}.} \bibinfo{year}{2017}\natexlab{}.
\newblock \showarticletitle{TREC 2017 Common Core Track Overview}. In \bibinfo{booktitle}{\emph{TREC}}.
\newblock


\bibitem[Askari et~al\mbox{.}(2023)]%
        {askari2023injecting}
\bibfield{author}{\bibinfo{person}{Arian Askari}, \bibinfo{person}{Amin Abolghasemi}, \bibinfo{person}{Gabriella Pasi}, \bibinfo{person}{Wessel Kraaij}, {and} \bibinfo{person}{Suzan Verberne}.} \bibinfo{year}{2023}\natexlab{}.
\newblock \showarticletitle{Injecting the BM25 Score as Text Improves BERT-Based Re-rankers}. In \bibinfo{booktitle}{\emph{Advances in Information Retrieval}}, \bibfield{editor}{\bibinfo{person}{Jaap Kamps}, \bibinfo{person}{Lorraine Goeuriot}, \bibinfo{person}{Fabio Crestani}, \bibinfo{person}{Maria Maistro}, \bibinfo{person}{Hideo Joho}, \bibinfo{person}{Brian Davis}, \bibinfo{person}{Cathal Gurrin}, \bibinfo{person}{Udo Kruschwitz}, {and} \bibinfo{person}{Annalina Caputo}} (Eds.). \bibinfo{publisher}{Springer Nature Switzerland}, \bibinfo{address}{Cham}, \bibinfo{pages}{66--83}.
\newblock
\showISBNx{978-3-031-28244-7}


\bibitem[Bajaj et~al\mbox{.}(2016)]%
        {bajaj2016ms}
\bibfield{author}{\bibinfo{person}{Payal Bajaj}, \bibinfo{person}{Daniel Campos}, \bibinfo{person}{Nick Craswell}, \bibinfo{person}{Li Deng}, \bibinfo{person}{Jianfeng Gao}, \bibinfo{person}{Xiaodong Liu}, \bibinfo{person}{Rangan Majumder}, \bibinfo{person}{Andrew McNamara}, \bibinfo{person}{Bhaskar Mitra}, \bibinfo{person}{Tri Nguyen}, {et~al\mbox{.}}} \bibinfo{year}{2016}\natexlab{}.
\newblock \showarticletitle{MS MARCO: A Human Generated Machine Reading Comprehension Dataset}.
\newblock \bibinfo{journal}{\emph{arXiv preprint arXiv:1611.09268}} (\bibinfo{year}{2016}).
\newblock


\bibitem[Chen et~al\mbox{.}(2023)]%
        {bge_m3}
\bibfield{author}{\bibinfo{person}{Jianlv Chen}, \bibinfo{person}{Shitao Xiao}, \bibinfo{person}{Peitian Zhang}, \bibinfo{person}{Kun Luo}, \bibinfo{person}{Defu Lian}, {and} \bibinfo{person}{Zheng Liu}.} \bibinfo{year}{2023}\natexlab{}.
\newblock \bibinfo{title}{BGE M3-Embedding: Multi-Lingual, Multi-Functionality, Multi-Granularity Text Embeddings Through Self-Knowledge Distillation}.
\newblock
\newblock
\showeprint[arxiv]{2309.07597}~[cs.CL]


\bibitem[Clark et~al\mbox{.}(2020)]%
        {clark2020electra}
\bibfield{author}{\bibinfo{person}{Kevin Clark}, \bibinfo{person}{Minh{-}Thang Luong}, \bibinfo{person}{Quoc~V. Le}, {and} \bibinfo{person}{Christopher~D. Manning}.} \bibinfo{year}{2020}\natexlab{}.
\newblock \showarticletitle{{ELECTRA:} Pre-training Text Encoders as Discriminators Rather Than Generators}.
\newblock \bibinfo{journal}{\emph{CoRR}}  \bibinfo{volume}{abs/2003.10555} (\bibinfo{year}{2020}).
\newblock
\showeprint[arXiv]{2003.10555}
\urldef\tempurl%
\url{https://arxiv.org/abs/2003.10555}
\showURL{%
\tempurl}


\bibitem[Craswell et~al\mbox{.}(2021)]%
        {craswell2021trecdl}
\bibfield{author}{\bibinfo{person}{Nick Craswell}, \bibinfo{person}{Bhaskar Mitra}, \bibinfo{person}{Emine Yilmaz}, \bibinfo{person}{Daniel Campos}, \bibinfo{person}{Ellen~M. Voorhees}, {and} \bibinfo{person}{Ian Soboroff}.} \bibinfo{year}{2021}\natexlab{}.
\newblock \showarticletitle{TREC Deep Learning Track: Reusable Test Collections in the Large Data Regime}. In \bibinfo{booktitle}{\emph{Proceedings of the 44th International ACM SIGIR Conference on Research and Development in Information Retrieval}} (Virtual Event, Canada) \emph{(\bibinfo{series}{SIGIR '21})}. \bibinfo{publisher}{Association for Computing Machinery}, \bibinfo{address}{New York, NY, USA}, \bibinfo{pages}{2369–2375}.
\newblock
\showISBNx{9781450380379}
\urldef\tempurl%
\url{https://doi.org/10.1145/3404835.3463249}
\showDOI{\tempurl}


\bibitem[Cronen-Townsend et~al\mbox{.}(2002)]%
        {cronen2002predicting}
\bibfield{author}{\bibinfo{person}{Steve Cronen-Townsend}, \bibinfo{person}{Yun Zhou}, {and} \bibinfo{person}{W.~Bruce Croft}.} \bibinfo{year}{2002}\natexlab{}.
\newblock \showarticletitle{Predicting query performance}. In \bibinfo{booktitle}{\emph{Proceedings of the 25th Annual International ACM SIGIR Conference on Research and Development in Information Retrieval}} (Tampere, Finland) \emph{(\bibinfo{series}{SIGIR '02})}. \bibinfo{publisher}{Association for Computing Machinery}, \bibinfo{address}{New York, NY, USA}, \bibinfo{pages}{299–306}.
\newblock
\showISBNx{1581135610}
\urldef\tempurl%
\url{https://doi.org/10.1145/564376.564429}
\showDOI{\tempurl}


\bibitem[Dai and Callan(2019)]%
        {dai2019deeper}
\bibfield{author}{\bibinfo{person}{Zhuyun Dai} {and} \bibinfo{person}{Jamie Callan}.} \bibinfo{year}{2019}\natexlab{}.
\newblock \showarticletitle{Deeper Text Understanding for {IR} with Contextual Neural Language Modeling}.
\newblock \bibinfo{journal}{\emph{CoRR}}  \bibinfo{volume}{abs/1905.09217} (\bibinfo{year}{2019}).
\newblock
\showeprint[arXiv]{1905.09217}
\urldef\tempurl%
\url{http://arxiv.org/abs/1905.09217}
\showURL{%
\tempurl}


\bibitem[Dai and Callan(2020)]%
        {dai2020context}
\bibfield{author}{\bibinfo{person}{Zhuyun Dai} {and} \bibinfo{person}{Jamie Callan}.} \bibinfo{year}{2020}\natexlab{}.
\newblock \showarticletitle{Context-Aware Term Weighting For First Stage Passage Retrieval}. In \bibinfo{booktitle}{\emph{Proceedings of the 43rd International ACM SIGIR Conference on Research and Development in Information Retrieval}} (Virtual Event, China) \emph{(\bibinfo{series}{SIGIR '20})}. \bibinfo{publisher}{Association for Computing Machinery}, \bibinfo{address}{New York, NY, USA}, \bibinfo{pages}{1533–1536}.
\newblock
\showISBNx{9781450380164}
\urldef\tempurl%
\url{https://doi.org/10.1145/3397271.3401204}
\showDOI{\tempurl}


\bibitem[Dai et~al\mbox{.}(2022)]%
        {dai2022promptagator}
\bibfield{author}{\bibinfo{person}{Zhuyun Dai}, \bibinfo{person}{Vincent~Y. Zhao}, \bibinfo{person}{Ji Ma}, \bibinfo{person}{Yi Luan}, \bibinfo{person}{Jianmo Ni}, \bibinfo{person}{Jing Lu}, \bibinfo{person}{Anton Bakalov}, \bibinfo{person}{Kelvin Guu}, \bibinfo{person}{Keith~B. Hall}, {and} \bibinfo{person}{Ming-Wei Chang}.} \bibinfo{year}{2022}\natexlab{}.
\newblock \bibinfo{title}{Promptagator: Few-shot Dense Retrieval From 8 Examples}.
\newblock
\newblock
\showeprint[arxiv]{2209.11755}~[cs.CL]
\urldef\tempurl%
\url{https://arxiv.org/abs/2209.11755}
\showURL{%
\tempurl}


\bibitem[Dalton et~al\mbox{.}(2014)]%
        {dalton2014entity}
\bibfield{author}{\bibinfo{person}{Jeffrey Dalton}, \bibinfo{person}{Laura Dietz}, {and} \bibinfo{person}{James Allan}.} \bibinfo{year}{2014}\natexlab{}.
\newblock \showarticletitle{Entity query feature expansion using knowledge base links}. In \bibinfo{booktitle}{\emph{Proceedings of the 37th international ACM SIGIR conference on Research \& development in information retrieval}}. ACM, \bibinfo{pages}{365--374}.
\newblock


\bibitem[Devlin et~al\mbox{.}(2018)]%
        {devlin2018bert}
\bibfield{author}{\bibinfo{person}{Jacob Devlin}, \bibinfo{person}{Ming{-}Wei Chang}, \bibinfo{person}{Kenton Lee}, {and} \bibinfo{person}{Kristina Toutanova}.} \bibinfo{year}{2018}\natexlab{}.
\newblock \showarticletitle{{BERT:} Pre-training of Deep Bidirectional Transformers for Language Understanding}.
\newblock \bibinfo{journal}{\emph{CoRR}}  \bibinfo{volume}{abs/1810.04805} (\bibinfo{year}{2018}).
\newblock
\showeprint[arXiv]{1810.04805}
\urldef\tempurl%
\url{http://arxiv.org/abs/1810.04805}
\showURL{%
\tempurl}


\bibitem[Dietz et~al\mbox{.}(2018)]%
        {dietz2018trec}
\bibfield{author}{\bibinfo{person}{Laura Dietz}, \bibinfo{person}{Ben Gamari}, \bibinfo{person}{Jeff Dalton}, {and} \bibinfo{person}{Nick Craswell}.} \bibinfo{year}{2018}\natexlab{}.
\newblock \showarticletitle{TREC Complex Answer Retrieval Overview.}. In \bibinfo{booktitle}{\emph{TREC}}.
\newblock


\bibitem[Dietz et~al\mbox{.}(2017)]%
        {dietz2017trec}
\bibfield{author}{\bibinfo{person}{Laura Dietz}, \bibinfo{person}{Manisha Verma}, \bibinfo{person}{Filip Radlinski}, {and} \bibinfo{person}{Nick Craswell}.} \bibinfo{year}{2017}\natexlab{}.
\newblock \showarticletitle{TREC Complex Answer Retrieval Overview.}. In \bibinfo{booktitle}{\emph{Proceedings of {Text REtrieval Conference} (TREC)}}.
\newblock


\bibitem[Ding et~al\mbox{.}(2024)]%
        {ding2024}
\bibfield{author}{\bibinfo{person}{Yifan Ding}, \bibinfo{person}{Amrit Poudel}, \bibinfo{person}{Qingkai Zeng}, \bibinfo{person}{Tim Weninger}, \bibinfo{person}{Balaji Veeramani}, {and} \bibinfo{person}{Sanmitra Bhattacharya}.} \bibinfo{year}{2024}\natexlab{}.
\newblock \bibinfo{title}{EntGPT: Linking Generative Large Language Models with Knowledge Bases}.
\newblock
\newblock
\showeprint[arxiv]{2402.06738}~[cs.CL]
\urldef\tempurl%
\url{https://arxiv.org/abs/2402.06738}
\showURL{%
\tempurl}


\bibitem[Ensan and Bagheri(2017)]%
        {ensan2017document}
\bibfield{author}{\bibinfo{person}{Faezeh Ensan} {and} \bibinfo{person}{Ebrahim Bagheri}.} \bibinfo{year}{2017}\natexlab{}.
\newblock \showarticletitle{Document Retrieval Model Through Semantic Linking}. In \bibinfo{booktitle}{\emph{Proceedings of the 10th ACM International Conference on Web Search and Data Mining}} (Cambridge, United Kingdom) \emph{(\bibinfo{series}{WSDM ’17})}. \bibinfo{publisher}{Association for Computing Machinery}, \bibinfo{address}{New York, NY, USA}, \bibinfo{pages}{181–190}.
\newblock
\showISBNx{9781450346757}
\urldef\tempurl%
\url{https://doi.org/10.1145/3018661.3018692}
\showDOI{\tempurl}


\bibitem[Formal et~al\mbox{.}(2021)]%
        {thibault2021splade}
\bibfield{author}{\bibinfo{person}{Thibault Formal}, \bibinfo{person}{Benjamin Piwowarski}, {and} \bibinfo{person}{St\'{e}phane Clinchant}.} \bibinfo{year}{2021}\natexlab{}.
\newblock \showarticletitle{SPLADE: Sparse Lexical and Expansion Model for First Stage Ranking}. In \bibinfo{booktitle}{\emph{Proceedings of the 44th International ACM SIGIR Conference on Research and Development in Information Retrieval}} (Virtual Event, Canada) \emph{(\bibinfo{series}{SIGIR '21})}. \bibinfo{publisher}{Association for Computing Machinery}, \bibinfo{address}{New York, NY, USA}, \bibinfo{pages}{2288–2292}.
\newblock
\showISBNx{9781450380379}
\urldef\tempurl%
\url{https://doi.org/10.1145/3404835.3463098}
\showDOI{\tempurl}


\bibitem[Gabrilovich and Markovitch(2009)]%
        {gabrilovich2009wikipedia}
\bibfield{author}{\bibinfo{person}{Evgeniy Gabrilovich} {and} \bibinfo{person}{Shaul Markovitch}.} \bibinfo{year}{2009}\natexlab{}.
\newblock \showarticletitle{Wikipedia-based Semantic Interpretation for Natural Language Processing}.
\newblock \bibinfo{journal}{\emph{Journal of Artificial Intelligence Research}}  \bibinfo{volume}{34} (\bibinfo{year}{2009}), \bibinfo{pages}{443--498}.
\newblock


\bibitem[Gao et~al\mbox{.}(2021)]%
        {gao2021complement}
\bibfield{author}{\bibinfo{person}{Luyu Gao}, \bibinfo{person}{Zhuyun Dai}, \bibinfo{person}{Tongfei Chen}, \bibinfo{person}{Zhen Fan}, \bibinfo{person}{Benjamin Van~Durme}, {and} \bibinfo{person}{Jamie Callan}.} \bibinfo{year}{2021}\natexlab{}.
\newblock \showarticletitle{Complement Lexical Retrieval Model with Semantic Residual Embeddings}. In \bibinfo{booktitle}{\emph{Advances in Information Retrieval: 43rd European Conference on IR Research, ECIR 2021, Virtual Event, March 28 – April 1, 2021, Proceedings, Part I}}. \bibinfo{publisher}{Springer-Verlag}, \bibinfo{address}{Berlin, Heidelberg}, \bibinfo{pages}{146–160}.
\newblock
\showISBNx{978-3-030-72112-1}
\urldef\tempurl%
\url{https://doi.org/10.1007/978-3-030-72113-8_10}
\showDOI{\tempurl}


\bibitem[Gao et~al\mbox{.}(2023)]%
        {gao-etal-2023-precise}
\bibfield{author}{\bibinfo{person}{Luyu Gao}, \bibinfo{person}{Xueguang Ma}, \bibinfo{person}{Jimmy Lin}, {and} \bibinfo{person}{Jamie Callan}.} \bibinfo{year}{2023}\natexlab{}.
\newblock \showarticletitle{Precise Zero-Shot Dense Retrieval without Relevance Labels}. In \bibinfo{booktitle}{\emph{Proceedings of the 61st Annual Meeting of the Association for Computational Linguistics (Volume 1: Long Papers)}}, \bibfield{editor}{\bibinfo{person}{Anna Rogers}, \bibinfo{person}{Jordan Boyd-Graber}, {and} \bibinfo{person}{Naoaki Okazaki}} (Eds.). \bibinfo{publisher}{Association for Computational Linguistics}, \bibinfo{address}{Toronto, Canada}, \bibinfo{pages}{1762--1777}.
\newblock
\urldef\tempurl%
\url{https://doi.org/10.18653/v1/2023.acl-long.99}
\showDOI{\tempurl}


\bibitem[Guo et~al\mbox{.}(2016)]%
        {guo2016drmm}
\bibfield{author}{\bibinfo{person}{Jiafeng Guo}, \bibinfo{person}{Yixing Fan}, \bibinfo{person}{Qingyao Ai}, {and} \bibinfo{person}{W.~Bruce Croft}.} \bibinfo{year}{2016}\natexlab{}.
\newblock \showarticletitle{A Deep Relevance Matching Model for Ad-Hoc Retrieval}. In \bibinfo{booktitle}{\emph{Proceedings of the 25th ACM International on Conference on Information and Knowledge Management}} (Indianapolis, Indiana, USA) \emph{(\bibinfo{series}{CIKM '16})}. \bibinfo{publisher}{Association for Computing Machinery}, \bibinfo{address}{New York, NY, USA}, \bibinfo{pages}{55–64}.
\newblock
\showISBNx{9781450340731}
\urldef\tempurl%
\url{https://doi.org/10.1145/2983323.2983769}
\showDOI{\tempurl}


\bibitem[He et~al\mbox{.}(2020)]%
        {he2020deberta}
\bibfield{author}{\bibinfo{person}{Pengcheng He}, \bibinfo{person}{Xiaodong Liu}, \bibinfo{person}{Jianfeng Gao}, {and} \bibinfo{person}{Weizhu Chen}.} \bibinfo{year}{2020}\natexlab{}.
\newblock \showarticletitle{DeBERTa: Decoding-enhanced {BERT} with Disentangled Attention}.
\newblock \bibinfo{journal}{\emph{CoRR}}  \bibinfo{volume}{abs/2006.03654} (\bibinfo{year}{2020}).
\newblock
\showeprint[arXiv]{2006.03654}
\urldef\tempurl%
\url{https://arxiv.org/abs/2006.03654}
\showURL{%
\tempurl}


\bibitem[Huang et~al\mbox{.}(2013)]%
        {huang2013learning}
\bibfield{author}{\bibinfo{person}{Po-Sen Huang}, \bibinfo{person}{Xiaodong He}, \bibinfo{person}{Jianfeng Gao}, \bibinfo{person}{Li Deng}, \bibinfo{person}{Alex Acero}, {and} \bibinfo{person}{Larry Heck}.} \bibinfo{year}{2013}\natexlab{}.
\newblock \showarticletitle{Learning Deep Structured Semantic Models for Web Search Using Clickthrough Data}. In \bibinfo{booktitle}{\emph{Proceedings of the 22nd ACM International Conference on Information \& Knowledge Management}} (San Francisco, California, USA) \emph{(\bibinfo{series}{CIKM '13})}. \bibinfo{publisher}{Association for Computing Machinery}, \bibinfo{address}{New York, NY, USA}, \bibinfo{pages}{2333–2338}.
\newblock
\showISBNx{9781450322638}
\urldef\tempurl%
\url{https://doi.org/10.1145/2505515.2505665}
\showDOI{\tempurl}


\bibitem[Hui et~al\mbox{.}(2017)]%
        {hui-etal-2017-pacrr}
\bibfield{author}{\bibinfo{person}{Kai Hui}, \bibinfo{person}{Andrew Yates}, \bibinfo{person}{Klaus Berberich}, {and} \bibinfo{person}{Gerard de Melo}.} \bibinfo{year}{2017}\natexlab{}.
\newblock \showarticletitle{{PACRR}: A Position-Aware Neural {IR} Model for Relevance Matching}. In \bibinfo{booktitle}{\emph{Proceedings of the 2017 Conference on Empirical Methods in Natural Language Processing}}. \bibinfo{publisher}{Association for Computational Linguistics}, \bibinfo{address}{Copenhagen, Denmark}, \bibinfo{pages}{1049--1058}.
\newblock
\urldef\tempurl%
\url{https://doi.org/10.18653/v1/D17-1110}
\showDOI{\tempurl}


\bibitem[Humeau et~al\mbox{.}(2020)]%
        {humeau2020polyencoders}
\bibfield{author}{\bibinfo{person}{Samuel Humeau}, \bibinfo{person}{Kurt Shuster}, \bibinfo{person}{Marie{-}Anne Lachaux}, {and} \bibinfo{person}{Jason Weston}.} \bibinfo{year}{2020}\natexlab{}.
\newblock \showarticletitle{Poly-encoders: Architectures and Pre-training Strategies for Fast and Accurate Multi-sentence Scoring}. In \bibinfo{booktitle}{\emph{8th International Conference on Learning Representations, {ICLR} 2020, Addis Ababa, Ethiopia, April 26-30, 2020}}. \bibinfo{publisher}{OpenReview.net}.
\newblock
\urldef\tempurl%
\url{https://openreview.net/forum?id=SkxgnnNFvH}
\showURL{%
\tempurl}


\bibitem[Jiang et~al\mbox{.}(2020)]%
        {jiang2020convbert}
\bibfield{author}{\bibinfo{person}{Zihang Jiang}, \bibinfo{person}{Weihao Yu}, \bibinfo{person}{Daquan Zhou}, \bibinfo{person}{Yunpeng Chen}, \bibinfo{person}{Jiashi Feng}, {and} \bibinfo{person}{Shuicheng Yan}.} \bibinfo{year}{2020}\natexlab{}.
\newblock \showarticletitle{ConvBERT: Improving {BERT} with Span-based Dynamic Convolution}.
\newblock \bibinfo{journal}{\emph{CoRR}}  \bibinfo{volume}{abs/2008.02496} (\bibinfo{year}{2020}).
\newblock
\showeprint[arXiv]{2008.02496}
\urldef\tempurl%
\url{https://arxiv.org/abs/2008.02496}
\showURL{%
\tempurl}


\bibitem[Jin et~al\mbox{.}(2025)]%
        {jin2025searchr1}
\bibfield{author}{\bibinfo{person}{Bowen Jin}, \bibinfo{person}{Hansi Zeng}, \bibinfo{person}{Zhenrui Yue}, \bibinfo{person}{Jinsung Yoon}, \bibinfo{person}{Sercan Arik}, \bibinfo{person}{Dong Wang}, \bibinfo{person}{Hamed Zamani}, {and} \bibinfo{person}{Jiawei Han}.} \bibinfo{year}{2025}\natexlab{}.
\newblock \bibinfo{title}{Search-R1: Training LLMs to Reason and Leverage Search Engines with Reinforcement Learning}.
\newblock
\newblock
\showeprint[arxiv]{2503.09516}~[cs.CL]
\urldef\tempurl%
\url{https://arxiv.org/abs/2503.09516}
\showURL{%
\tempurl}


\bibitem[Kamphuis et~al\mbox{.}(2023a)]%
        {kamphuis2023}
\bibfield{author}{\bibinfo{person}{Chris Kamphuis}, \bibinfo{person}{Aileen Lin}, \bibinfo{person}{Siwen Yang}, \bibinfo{person}{Jimmy Lin}, \bibinfo{person}{Arjen~P. de Vries}, {and} \bibinfo{person}{Faegheh Hasibi}.} \bibinfo{year}{2023}\natexlab{a}.
\newblock \showarticletitle{MMEAD: MS MARCO Entity Annotations and Disambiguations}. In \bibinfo{booktitle}{\emph{Proceedings of the 46th International ACM SIGIR Conference on Research and Development in Information Retrieval}} (Taipei, Taiwan) \emph{(\bibinfo{series}{SIGIR '23})}. \bibinfo{publisher}{Association for Computing Machinery}, \bibinfo{address}{New York, NY, USA}, \bibinfo{pages}{2817–2825}.
\newblock
\showISBNx{9781450394086}
\urldef\tempurl%
\url{https://doi.org/10.1145/3539618.3591887}
\showDOI{\tempurl}


\bibitem[Kamphuis et~al\mbox{.}(2023b)]%
        {Kamphuis2023mmead}
\bibfield{author}{\bibinfo{person}{Chris Kamphuis}, \bibinfo{person}{Aileen Lin}, \bibinfo{person}{Siwen Yang}, \bibinfo{person}{Jimmy Lin}, \bibinfo{person}{Arjen~P. de Vries}, {and} \bibinfo{person}{Faegheh Hasibi}.} \bibinfo{year}{2023}\natexlab{b}.
\newblock \showarticletitle{MMEAD: MS MARCO Entity Annotations and Disambiguations}. In \bibinfo{booktitle}{\emph{Proceedings of the 46th International ACM SIGIR Conference on Research and Development in Information Retrieval}} (Taipei, Taiwan) \emph{(\bibinfo{series}{SIGIR '23})}. \bibinfo{publisher}{Association for Computing Machinery}, \bibinfo{address}{New York, NY, USA}, \bibinfo{pages}{2817–2825}.
\newblock
\showISBNx{9781450394086}
\urldef\tempurl%
\url{https://doi.org/10.1145/3539618.3591887}
\showDOI{\tempurl}


\bibitem[Karpukhin et~al\mbox{.}(2020)]%
        {karpukhin-etal-2020-dense}
\bibfield{author}{\bibinfo{person}{Vladimir Karpukhin}, \bibinfo{person}{Barlas Oguz}, \bibinfo{person}{Sewon Min}, \bibinfo{person}{Patrick Lewis}, \bibinfo{person}{Ledell Wu}, \bibinfo{person}{Sergey Edunov}, \bibinfo{person}{Danqi Chen}, {and} \bibinfo{person}{Wen-tau Yih}.} \bibinfo{year}{2020}\natexlab{}.
\newblock \showarticletitle{Dense Passage Retrieval for Open-Domain Question Answering}. In \bibinfo{booktitle}{\emph{Proceedings of the 2020 Conference on Empirical Methods in Natural Language Processing (EMNLP)}}. \bibinfo{publisher}{Association for Computational Linguistics}, \bibinfo{address}{Online}, \bibinfo{pages}{6769--6781}.
\newblock
\urldef\tempurl%
\url{https://doi.org/10.18653/v1/2020.emnlp-main.550}
\showDOI{\tempurl}


\bibitem[Khattab and Zaharia(2020)]%
        {khattab2020colbert}
\bibfield{author}{\bibinfo{person}{Omar Khattab} {and} \bibinfo{person}{Matei Zaharia}.} \bibinfo{year}{2020}\natexlab{}.
\newblock \showarticletitle{ColBERT: Efficient and Effective Passage Search via Contextualized Late Interaction over BERT}. In \bibinfo{booktitle}{\emph{Proceedings of the 43rd International ACM SIGIR Conference on Research and Development in Information Retrieval}} (Virtual Event, China) \emph{(\bibinfo{series}{SIGIR '20})}. \bibinfo{publisher}{Association for Computing Machinery}, \bibinfo{address}{New York, NY, USA}, \bibinfo{pages}{39–48}.
\newblock
\showISBNx{9781450380164}
\urldef\tempurl%
\url{https://doi.org/10.1145/3397271.3401075}
\showDOI{\tempurl}


\bibitem[Kingma and Ba(2014)]%
        {kingma2014adam}
\bibfield{author}{\bibinfo{person}{Diederik~P Kingma} {and} \bibinfo{person}{Jimmy Ba}.} \bibinfo{year}{2014}\natexlab{}.
\newblock \showarticletitle{Adam: A Method for Stochastic Optimization}.
\newblock \bibinfo{journal}{\emph{arXiv preprint arXiv:1412.6980}} (\bibinfo{year}{2014}).
\newblock


\bibitem[Lassance et~al\mbox{.}(2024)]%
        {lassance2024spladev3}
\bibfield{author}{\bibinfo{person}{Carlos Lassance}, \bibinfo{person}{Hervé Déjean}, \bibinfo{person}{Thibault Formal}, {and} \bibinfo{person}{Stéphane Clinchant}.} \bibinfo{year}{2024}\natexlab{}.
\newblock \bibinfo{title}{SPLADE-v3: New Baselines for SPLADE}.
\newblock
\newblock
\showeprint[arxiv]{2403.06789}~[cs.IR]
\urldef\tempurl%
\url{https://arxiv.org/abs/2403.06789}
\showURL{%
\tempurl}


\bibitem[Li et~al\mbox{.}(2018)]%
        {li-etal-2018-nprf}
\bibfield{author}{\bibinfo{person}{Canjia Li}, \bibinfo{person}{Yingfei Sun}, \bibinfo{person}{Ben He}, \bibinfo{person}{Le Wang}, \bibinfo{person}{Kai Hui}, \bibinfo{person}{Andrew Yates}, \bibinfo{person}{Le Sun}, {and} \bibinfo{person}{Jungang Xu}.} \bibinfo{year}{2018}\natexlab{}.
\newblock \showarticletitle{{NPRF}: A Neural Pseudo Relevance Feedback Framework for Ad-hoc Information Retrieval}. In \bibinfo{booktitle}{\emph{Proceedings of the 2018 Conference on Empirical Methods in Natural Language Processing}}. \bibinfo{publisher}{Association for Computational Linguistics}, \bibinfo{address}{Brussels, Belgium}, \bibinfo{pages}{4482--4491}.
\newblock
\urldef\tempurl%
\url{https://doi.org/10.18653/v1/D18-1478}
\showDOI{\tempurl}


\bibitem[Li et~al\mbox{.}(2020)]%
        {li2020parade}
\bibfield{author}{\bibinfo{person}{Canjia Li}, \bibinfo{person}{Andrew Yates}, \bibinfo{person}{Sean MacAvaney}, \bibinfo{person}{Ben He}, {and} \bibinfo{person}{Yingfei Sun}.} \bibinfo{year}{2020}\natexlab{}.
\newblock \showarticletitle{{PARADE:} Passage Representation Aggregation for Document Reranking}.
\newblock \bibinfo{journal}{\emph{CoRR}}  \bibinfo{volume}{abs/2008.09093} (\bibinfo{year}{2020}).
\newblock
\showeprint[arXiv]{2008.09093}
\urldef\tempurl%
\url{https://arxiv.org/abs/2008.09093}
\showURL{%
\tempurl}


\bibitem[Lin et~al\mbox{.}(2020)]%
        {lin2020distilling}
\bibfield{author}{\bibinfo{person}{Sheng{-}Chieh Lin}, \bibinfo{person}{Jheng{-}Hong Yang}, {and} \bibinfo{person}{Jimmy Lin}.} \bibinfo{year}{2020}\natexlab{}.
\newblock \showarticletitle{Distilling Dense Representations for Ranking using Tightly-Coupled Teachers}.
\newblock \bibinfo{journal}{\emph{CoRR}}  \bibinfo{volume}{abs/2010.11386} (\bibinfo{year}{2020}).
\newblock
\showeprint[arXiv]{2010.11386}
\urldef\tempurl%
\url{https://arxiv.org/abs/2010.11386}
\showURL{%
\tempurl}


\bibitem[Lin et~al\mbox{.}(2021)]%
        {lin-etal-2021-batch}
\bibfield{author}{\bibinfo{person}{Sheng-Chieh Lin}, \bibinfo{person}{Jheng-Hong Yang}, {and} \bibinfo{person}{Jimmy Lin}.} \bibinfo{year}{2021}\natexlab{}.
\newblock \showarticletitle{In-Batch Negatives for Knowledge Distillation with Tightly-Coupled Teachers for Dense Retrieval}. In \bibinfo{booktitle}{\emph{Proceedings of the 6th Workshop on Representation Learning for NLP (RepL4NLP-2021)}}, \bibfield{editor}{\bibinfo{person}{Anna Rogers}, \bibinfo{person}{Iacer Calixto}, \bibinfo{person}{Ivan Vuli{\'c}}, \bibinfo{person}{Naomi Saphra}, \bibinfo{person}{Nora Kassner}, \bibinfo{person}{Oana-Maria Camburu}, \bibinfo{person}{Trapit Bansal}, {and} \bibinfo{person}{Vered Shwartz}} (Eds.). \bibinfo{publisher}{Association for Computational Linguistics}, \bibinfo{address}{Online}, \bibinfo{pages}{163--173}.
\newblock
\urldef\tempurl%
\url{https://doi.org/10.18653/v1/2021.repl4nlp-1.17}
\showDOI{\tempurl}


\bibitem[Liu and Fang(2015)]%
        {liu2015latent}
\bibfield{author}{\bibinfo{person}{Xitong Liu} {and} \bibinfo{person}{Hui Fang}.} \bibinfo{year}{2015}\natexlab{}.
\newblock \showarticletitle{Latent entity space: a novel retrieval approach for entity-bearing queries}.
\newblock \bibinfo{journal}{\emph{Information Retrieval Journal}} \bibinfo{volume}{18}, \bibinfo{number}{6} (\bibinfo{year}{2015}), \bibinfo{pages}{473--503}.
\newblock


\bibitem[Liu et~al\mbox{.}(2019)]%
        {liu2019roberta}
\bibfield{author}{\bibinfo{person}{Yinhan Liu}, \bibinfo{person}{Myle Ott}, \bibinfo{person}{Naman Goyal}, \bibinfo{person}{Jingfei Du}, \bibinfo{person}{Mandar Joshi}, \bibinfo{person}{Danqi Chen}, \bibinfo{person}{Omer Levy}, \bibinfo{person}{Mike Lewis}, \bibinfo{person}{Luke Zettlemoyer}, {and} \bibinfo{person}{Veselin Stoyanov}.} \bibinfo{year}{2019}\natexlab{}.
\newblock \showarticletitle{RoBERTa: {A} Robustly Optimized {BERT} Pretraining Approach}.
\newblock \bibinfo{journal}{\emph{CoRR}}  \bibinfo{volume}{abs/1907.11692} (\bibinfo{year}{2019}).
\newblock
\showeprint[arXiv]{1907.11692}
\urldef\tempurl%
\url{http://arxiv.org/abs/1907.11692}
\showURL{%
\tempurl}


\bibitem[Liu et~al\mbox{.}(2018)]%
        {liu-etal-2018-entity}
\bibfield{author}{\bibinfo{person}{Zhenghao Liu}, \bibinfo{person}{Chenyan Xiong}, \bibinfo{person}{Maosong Sun}, {and} \bibinfo{person}{Zhiyuan Liu}.} \bibinfo{year}{2018}\natexlab{}.
\newblock \showarticletitle{Entity-Duet Neural Ranking: Understanding the Role of Knowledge Graph Semantics in Neural Information Retrieval}. In \bibinfo{booktitle}{\emph{Proceedings of the 56th Annual Meeting of the Association for Computational Linguistics (Volume 1: Long Papers)}}. \bibinfo{publisher}{Association for Computational Linguistics}, \bibinfo{address}{Melbourne, Australia}, \bibinfo{pages}{2395--2405}.
\newblock
\urldef\tempurl%
\url{https://doi.org/10.18653/v1/P18-1223}
\showDOI{\tempurl}


\bibitem[Ma et~al\mbox{.}(2021)]%
        {ma2021a}
\bibfield{author}{\bibinfo{person}{Xueguang Ma}, \bibinfo{person}{Kai Sun}, \bibinfo{person}{Ronak Pradeep}, {and} \bibinfo{person}{Jimmy Lin}.} \bibinfo{year}{2021}\natexlab{}.
\newblock \showarticletitle{A Replication Study of Dense Passage Retriever}.
\newblock \bibinfo{journal}{\emph{CoRR}}  \bibinfo{volume}{abs/2104.05740} (\bibinfo{year}{2021}).
\newblock
\showeprint[arXiv]{2104.05740}
\urldef\tempurl%
\url{https://arxiv.org/abs/2104.05740}
\showURL{%
\tempurl}


\bibitem[Mackie et~al\mbox{.}(2022)]%
        {mackie2022codec}
\bibfield{author}{\bibinfo{person}{Iain Mackie}, \bibinfo{person}{Paul Owoicho}, \bibinfo{person}{Carlos Gemmell}, \bibinfo{person}{Sophie Fischer}, \bibinfo{person}{Sean MacAvaney}, {and} \bibinfo{person}{Jeffrey Dalton}.} \bibinfo{year}{2022}\natexlab{}.
\newblock \showarticletitle{CODEC: Complex Document and Entity Collection}. In \bibinfo{booktitle}{\emph{Proceedings of the 45th International ACM SIGIR Conference on Research and Development in Information Retrieval}} (Madrid, Spain) \emph{(\bibinfo{series}{SIGIR '22})}. \bibinfo{publisher}{Association for Computing Machinery}, \bibinfo{address}{New York, NY, USA}, \bibinfo{pages}{3067–3077}.
\newblock
\showISBNx{9781450387323}
\urldef\tempurl%
\url{https://doi.org/10.1145/3477495.3531712}
\showDOI{\tempurl}


\bibitem[Mitra and Craswell(2019)]%
        {mitra2019an}
\bibfield{author}{\bibinfo{person}{Bhaskar Mitra} {and} \bibinfo{person}{Nick Craswell}.} \bibinfo{year}{2019}\natexlab{}.
\newblock \showarticletitle{An Updated Duet Model for Passage Re-ranking}.
\newblock \bibinfo{journal}{\emph{CoRR}}  \bibinfo{volume}{abs/1903.07666} (\bibinfo{year}{2019}).
\newblock
\showeprint[arXiv]{1903.07666}
\urldef\tempurl%
\url{http://arxiv.org/abs/1903.07666}
\showURL{%
\tempurl}


\bibitem[Naseri et~al\mbox{.}(2021)]%
        {naseri2021ceqe}
\bibfield{author}{\bibinfo{person}{Shahrzad Naseri}, \bibinfo{person}{Jeffrey Dalton}, \bibinfo{person}{Andrew Yates}, {and} \bibinfo{person}{James Allan}.} \bibinfo{year}{2021}\natexlab{}.
\newblock \showarticletitle{CEQE: Contextualized Embeddings for Query Expansion}. In \bibinfo{booktitle}{\emph{Advances in Information Retrieval: 43rd European Conference on IR Research, ECIR 2021, Virtual Event, March 28 – April 1, 2021, Proceedings, Part I}}. \bibinfo{publisher}{Springer-Verlag}, \bibinfo{address}{Berlin, Heidelberg}, \bibinfo{pages}{467–482}.
\newblock
\showISBNx{978-3-030-72112-1}
\urldef\tempurl%
\url{https://doi.org/10.1007/978-3-030-72113-8\_31}
\showDOI{\tempurl}


\bibitem[Nguyen et~al\mbox{.}(2024)]%
        {nguyen2024dyvo}
\bibfield{author}{\bibinfo{person}{Thong Nguyen}, \bibinfo{person}{Shubham Chatterjee}, \bibinfo{person}{Sean MacAvaney}, \bibinfo{person}{Iain Mackie}, \bibinfo{person}{Jeff Dalton}, {and} \bibinfo{person}{Andrew Yates}.} \bibinfo{year}{2024}\natexlab{}.
\newblock \showarticletitle{{D}y{V}o: Dynamic Vocabularies for Learned Sparse Retrieval with Entities}. In \bibinfo{booktitle}{\emph{Proceedings of the 2024 Conference on Empirical Methods in Natural Language Processing}}, \bibfield{editor}{\bibinfo{person}{Yaser Al-Onaizan}, \bibinfo{person}{Mohit Bansal}, {and} \bibinfo{person}{Yun-Nung Chen}} (Eds.). \bibinfo{publisher}{Association for Computational Linguistics}, \bibinfo{address}{Miami, Florida, USA}, \bibinfo{pages}{767--783}.
\newblock
\urldef\tempurl%
\url{https://doi.org/10.18653/v1/2024.emnlp-main.45}
\showDOI{\tempurl}


\bibitem[Nguyen et~al\mbox{.}(2016)]%
        {nguyen2016marco}
\bibfield{author}{\bibinfo{person}{Tri Nguyen}, \bibinfo{person}{Mir Rosenberg}, \bibinfo{person}{Xia Song}, \bibinfo{person}{Jianfeng Gao}, \bibinfo{person}{Saurabh Tiwary}, \bibinfo{person}{Rangan Majumder}, {and} \bibinfo{person}{Li Deng}.} \bibinfo{year}{2016}\natexlab{}.
\newblock \showarticletitle{{MS} {MARCO:} {A} Human Generated MAchine Reading COmprehension Dataset}.
\newblock \bibinfo{journal}{\emph{CoRR}}  \bibinfo{volume}{abs/1611.09268} (\bibinfo{year}{2016}).
\newblock
\showeprint[arXiv]{1611.09268}
\urldef\tempurl%
\url{http://arxiv.org/abs/1611.09268}
\showURL{%
\tempurl}


\bibitem[Nogueira and Cho(2019)]%
        {nogueira2019passage}
\bibfield{author}{\bibinfo{person}{Rodrigo~Frassetto Nogueira} {and} \bibinfo{person}{Kyunghyun Cho}.} \bibinfo{year}{2019}\natexlab{}.
\newblock \showarticletitle{Passage Re-ranking with {BERT}}.
\newblock \bibinfo{journal}{\emph{CoRR}}  \bibinfo{volume}{abs/1901.04085} (\bibinfo{year}{2019}).
\newblock
\showeprint[arXiv]{1901.04085}
\urldef\tempurl%
\url{http://arxiv.org/abs/1901.04085}
\showURL{%
\tempurl}


\bibitem[Nogueira et~al\mbox{.}(2019)]%
        {nogueira2019document}
\bibfield{author}{\bibinfo{person}{Rodrigo~Frassetto Nogueira}, \bibinfo{person}{Wei Yang}, \bibinfo{person}{Jimmy Lin}, {and} \bibinfo{person}{Kyunghyun Cho}.} \bibinfo{year}{2019}\natexlab{}.
\newblock \showarticletitle{Document Expansion by Query Prediction}.
\newblock \bibinfo{journal}{\emph{CoRR}}  \bibinfo{volume}{abs/1904.08375} (\bibinfo{year}{2019}).
\newblock
\showeprint[arXiv]{1904.08375}
\urldef\tempurl%
\url{http://arxiv.org/abs/1904.08375}
\showURL{%
\tempurl}


\bibitem[Piccinno and Ferragina(2014)]%
        {piccinno2014wat}
\bibfield{author}{\bibinfo{person}{Francesco Piccinno} {and} \bibinfo{person}{Paolo Ferragina}.} \bibinfo{year}{2014}\natexlab{}.
\newblock \showarticletitle{From TagME to WAT: A New Entity Annotator}. In \bibinfo{booktitle}{\emph{Proceedings of the First International Workshop on Entity Recognition \& Disambiguation}} (Gold Coast, Queensland, Australia) \emph{(\bibinfo{series}{ERD '14})}. \bibinfo{publisher}{Association for Computing Machinery}, \bibinfo{address}{New York, NY, USA}, \bibinfo{pages}{55–62}.
\newblock
\showISBNx{9781450330237}
\urldef\tempurl%
\url{https://doi.org/10.1145/2633211.2634350}
\showDOI{\tempurl}


\bibitem[Pradeep et~al\mbox{.}(2023a)]%
        {pradeep2023rankvicuna}
\bibfield{author}{\bibinfo{person}{Ronak Pradeep}, \bibinfo{person}{Sahel Sharifymoghaddam}, {and} \bibinfo{person}{Jimmy Lin}.} \bibinfo{year}{2023}\natexlab{a}.
\newblock \bibinfo{title}{RankVicuna: Zero-Shot Listwise Document Reranking with Open-Source Large Language Models}.
\newblock
\newblock
\showeprint[arxiv]{2309.15088}~[cs.IR]
\urldef\tempurl%
\url{https://arxiv.org/abs/2309.15088}
\showURL{%
\tempurl}


\bibitem[Pradeep et~al\mbox{.}(2023b)]%
        {pradeep2023rankzephyr}
\bibfield{author}{\bibinfo{person}{Ronak Pradeep}, \bibinfo{person}{Sahel Sharifymoghaddam}, {and} \bibinfo{person}{Jimmy Lin}.} \bibinfo{year}{2023}\natexlab{b}.
\newblock \bibinfo{title}{RankZephyr: Effective and Robust Zero-Shot Listwise Reranking is a Breeze!}
\newblock
\newblock
\showeprint[arxiv]{2312.02724}~[cs.IR]
\urldef\tempurl%
\url{https://arxiv.org/abs/2312.02724}
\showURL{%
\tempurl}


\bibitem[Raviv et~al\mbox{.}(2016)]%
        {raviv2016document}
\bibfield{author}{\bibinfo{person}{Hadas Raviv}, \bibinfo{person}{Oren Kurland}, {and} \bibinfo{person}{David Carmel}.} \bibinfo{year}{2016}\natexlab{}.
\newblock \showarticletitle{Document Retrieval Using Entity-Based Language Models}. In \bibinfo{booktitle}{\emph{Proceedings of the 39th International ACM SIGIR Conference on Research and Development in Information Retrieval}} (Pisa, Italy) \emph{(\bibinfo{series}{SIGIR ’16})}. \bibinfo{publisher}{Association for Computing Machinery}, \bibinfo{address}{New York, NY, USA}, \bibinfo{pages}{65–74}.
\newblock
\showISBNx{9781450340694}
\urldef\tempurl%
\url{https://doi.org/10.1145/2911451.2911508}
\showDOI{\tempurl}


\bibitem[Reimers and Gurevych(2019)]%
        {reimers2019sentence}
\bibfield{author}{\bibinfo{person}{Nils Reimers} {and} \bibinfo{person}{Iryna Gurevych}.} \bibinfo{year}{2019}\natexlab{}.
\newblock \showarticletitle{Sentence-BERT: Sentence Embeddings using Siamese BERT-Networks}.
\newblock \bibinfo{journal}{\emph{CoRR}}  \bibinfo{volume}{abs/1908.10084} (\bibinfo{year}{2019}).
\newblock
\showeprint[arXiv]{1908.10084}
\urldef\tempurl%
\url{http://arxiv.org/abs/1908.10084}
\showURL{%
\tempurl}


\bibitem[Shen et~al\mbox{.}(2023)]%
        {shen2023lexmae}
\bibfield{author}{\bibinfo{person}{Tao Shen}, \bibinfo{person}{Xiubo Geng}, \bibinfo{person}{Chongyang Tao}, \bibinfo{person}{Can Xu}, \bibinfo{person}{Xiaolong Huang}, \bibinfo{person}{Binxing Jiao}, \bibinfo{person}{Linjun Yang}, {and} \bibinfo{person}{Daxin Jiang}.} \bibinfo{year}{2023}\natexlab{}.
\newblock \bibinfo{title}{LexMAE: Lexicon-Bottlenecked Pretraining for Large-Scale Retrieval}.
\newblock
\newblock
\showeprint[arxiv]{2208.14754}~[cs.IR]
\urldef\tempurl%
\url{https://arxiv.org/abs/2208.14754}
\showURL{%
\tempurl}


\bibitem[Shen et~al\mbox{.}(2015)]%
        {shen2015entity}
\bibfield{author}{\bibinfo{person}{Wei Shen}, \bibinfo{person}{Jianyong Wang}, {and} \bibinfo{person}{Jiawei Han}.} \bibinfo{year}{2015}\natexlab{}.
\newblock \showarticletitle{Entity Linking with a Knowledge Base: Issues, Techniques, and Solutions}.
\newblock \bibinfo{journal}{\emph{IEEE Transactions on Knowledge and Data Engineering}} \bibinfo{volume}{27}, \bibinfo{number}{2} (\bibinfo{year}{2015}), \bibinfo{pages}{443--460}.
\newblock
\urldef\tempurl%
\url{https://doi.org/10.1109/TKDE.2014.2327028}
\showDOI{\tempurl}


\bibitem[Su et~al\mbox{.}(2023)]%
        {su-etal-2023-one}
\bibfield{author}{\bibinfo{person}{Hongjin Su}, \bibinfo{person}{Weijia Shi}, \bibinfo{person}{Jungo Kasai}, \bibinfo{person}{Yizhong Wang}, \bibinfo{person}{Yushi Hu}, \bibinfo{person}{Mari Ostendorf}, \bibinfo{person}{Wen-tau Yih}, \bibinfo{person}{Noah~A. Smith}, \bibinfo{person}{Luke Zettlemoyer}, {and} \bibinfo{person}{Tao Yu}.} \bibinfo{year}{2023}\natexlab{}.
\newblock \showarticletitle{One Embedder, Any Task: Instruction-Finetuned Text Embeddings}. In \bibinfo{booktitle}{\emph{Findings of the Association for Computational Linguistics: ACL 2023}}, \bibfield{editor}{\bibinfo{person}{Anna Rogers}, \bibinfo{person}{Jordan Boyd-Graber}, {and} \bibinfo{person}{Naoaki Okazaki}} (Eds.). \bibinfo{publisher}{Association for Computational Linguistics}, \bibinfo{address}{Toronto, Canada}, \bibinfo{pages}{1102--1121}.
\newblock
\urldef\tempurl%
\url{https://doi.org/10.18653/v1/2023.findings-acl.71}
\showDOI{\tempurl}


\bibitem[Vollmers et~al\mbox{.}(2025)]%
        {vollmers-etal-2025-contextual}
\bibfield{author}{\bibinfo{person}{Daniel Vollmers}, \bibinfo{person}{Hamada Zahera}, \bibinfo{person}{Diego Moussallem}, {and} \bibinfo{person}{Axel-Cyrille Ngonga~Ngomo}.} \bibinfo{year}{2025}\natexlab{}.
\newblock \showarticletitle{Contextual Augmentation for Entity Linking using Large Language Models}. In \bibinfo{booktitle}{\emph{Proceedings of the 31st International Conference on Computational Linguistics}}, \bibfield{editor}{\bibinfo{person}{Owen Rambow}, \bibinfo{person}{Leo Wanner}, \bibinfo{person}{Marianna Apidianaki}, \bibinfo{person}{Hend Al-Khalifa}, \bibinfo{person}{Barbara~Di Eugenio}, {and} \bibinfo{person}{Steven Schockaert}} (Eds.). \bibinfo{publisher}{Association for Computational Linguistics}, \bibinfo{address}{Abu Dhabi, UAE}, \bibinfo{pages}{8535--8545}.
\newblock
\urldef\tempurl%
\url{https://aclanthology.org/2025.coling-main.570/}
\showURL{%
\tempurl}


\bibitem[Voorhees et~al\mbox{.}(2003)]%
        {voorhees2003overview}
\bibfield{author}{\bibinfo{person}{Ellen~M Voorhees} {et~al\mbox{.}}} \bibinfo{year}{2003}\natexlab{}.
\newblock \showarticletitle{Overview of the TREC 2003 robust retrieval track.}. In \bibinfo{booktitle}{\emph{Trec}}. \bibinfo{pages}{69--77}.
\newblock


\bibitem[Wang et~al\mbox{.}(2024)]%
        {wang2024textembeddings}
\bibfield{author}{\bibinfo{person}{Liang Wang}, \bibinfo{person}{Nan Yang}, \bibinfo{person}{Xiaolong Huang}, \bibinfo{person}{Binxing Jiao}, \bibinfo{person}{Linjun Yang}, \bibinfo{person}{Daxin Jiang}, \bibinfo{person}{Rangan Majumder}, {and} \bibinfo{person}{Furu Wei}.} \bibinfo{year}{2024}\natexlab{}.
\newblock \bibinfo{title}{Text Embeddings by Weakly-Supervised Contrastive Pre-training}.
\newblock
\newblock
\showeprint[arxiv]{2212.03533}~[cs.CL]
\urldef\tempurl%
\url{https://arxiv.org/abs/2212.03533}
\showURL{%
\tempurl}


\bibitem[Wang et~al\mbox{.}(2021)]%
        {wang2021bert}
\bibfield{author}{\bibinfo{person}{Shuai Wang}, \bibinfo{person}{Shengyao Zhuang}, {and} \bibinfo{person}{Guido Zuccon}.} \bibinfo{year}{2021}\natexlab{}.
\newblock \showarticletitle{BERT-based Dense Retrievers Require Interpolation with BM25 for Effective Passage Retrieval}. In \bibinfo{booktitle}{\emph{Proceedings of the 2021 ACM SIGIR International Conference on Theory of Information Retrieval}} (Virtual Event, Canada) \emph{(\bibinfo{series}{ICTIR '21})}. \bibinfo{publisher}{Association for Computing Machinery}, \bibinfo{address}{New York, NY, USA}, \bibinfo{pages}{317–324}.
\newblock
\showISBNx{9781450386111}
\urldef\tempurl%
\url{https://doi.org/10.1145/3471158.3472233}
\showDOI{\tempurl}


\bibitem[Wang et~al\mbox{.}(2023)]%
        {xiao2023colbert-prf}
\bibfield{author}{\bibinfo{person}{Xiao Wang}, \bibinfo{person}{Craig MacDonald}, \bibinfo{person}{Nicola Tonellotto}, {and} \bibinfo{person}{Iadh Ounis}.} \bibinfo{year}{2023}\natexlab{}.
\newblock \showarticletitle{ColBERT-PRF: Semantic Pseudo-Relevance Feedback for Dense Passage and Document Retrieval}.
\newblock \bibinfo{journal}{\emph{ACM Trans. Web}} \bibinfo{volume}{17}, \bibinfo{number}{1}, Article \bibinfo{articleno}{3} (\bibinfo{date}{jan} \bibinfo{year}{2023}), \bibinfo{numpages}{39}~pages.
\newblock
\showISSN{1559-1131}
\urldef\tempurl%
\url{https://doi.org/10.1145/3572405}
\showDOI{\tempurl}


\bibitem[Xiao et~al\mbox{.}(2022)]%
        {xiao-etal-2022-retromae}
\bibfield{author}{\bibinfo{person}{Shitao Xiao}, \bibinfo{person}{Zheng Liu}, \bibinfo{person}{Yingxia Shao}, {and} \bibinfo{person}{Zhao Cao}.} \bibinfo{year}{2022}\natexlab{}.
\newblock \showarticletitle{{R}etro{MAE}: Pre-Training Retrieval-oriented Language Models Via Masked Auto-Encoder}. In \bibinfo{booktitle}{\emph{Proceedings of the 2022 Conference on Empirical Methods in Natural Language Processing}}, \bibfield{editor}{\bibinfo{person}{Yoav Goldberg}, \bibinfo{person}{Zornitsa Kozareva}, {and} \bibinfo{person}{Yue Zhang}} (Eds.). \bibinfo{publisher}{Association for Computational Linguistics}, \bibinfo{address}{Abu Dhabi, United Arab Emirates}, \bibinfo{pages}{538--548}.
\newblock
\urldef\tempurl%
\url{https://doi.org/10.18653/v1/2022.emnlp-main.35}
\showDOI{\tempurl}


\bibitem[Xin et~al\mbox{.}(2024)]%
        {xin2024}
\bibfield{author}{\bibinfo{person}{Amy Xin}, \bibinfo{person}{Yunjia Qi}, \bibinfo{person}{Zijun Yao}, \bibinfo{person}{Fangwei Zhu}, \bibinfo{person}{Kaisheng Zeng}, \bibinfo{person}{Xu Bin}, \bibinfo{person}{Lei Hou}, {and} \bibinfo{person}{Juanzi Li}.} \bibinfo{year}{2024}\natexlab{}.
\newblock \bibinfo{title}{LLMAEL: Large Language Models are Good Context Augmenters for Entity Linking}.
\newblock
\newblock
\showeprint[arxiv]{2407.04020}~[cs.CL]
\urldef\tempurl%
\url{https://arxiv.org/abs/2407.04020}
\showURL{%
\tempurl}


\bibitem[Xiong et~al\mbox{.}(2017a)]%
        {xiong2017word}
\bibfield{author}{\bibinfo{person}{Chenyan Xiong}, \bibinfo{person}{Jamie Callan}, {and} \bibinfo{person}{Tie-Yan Liu}.} \bibinfo{year}{2017}\natexlab{a}.
\newblock \showarticletitle{Word-Entity Duet Representations for Document Ranking}. In \bibinfo{booktitle}{\emph{Proceedings of the 40th International ACM SIGIR Conference on Research and Development in Information Retrieval}} (Shinjuku, Tokyo, Japan) \emph{(\bibinfo{series}{SIGIR ’17})}. \bibinfo{publisher}{Association for Computing Machinery}, \bibinfo{address}{New York, NY, USA}, \bibinfo{pages}{763–772}.
\newblock
\showISBNx{9781450350228}
\urldef\tempurl%
\url{https://doi.org/10.1145/3077136.3080768}
\showDOI{\tempurl}


\bibitem[Xiong et~al\mbox{.}(2017b)]%
        {xiong2017end}
\bibfield{author}{\bibinfo{person}{Chenyan Xiong}, \bibinfo{person}{Zhuyun Dai}, \bibinfo{person}{Jamie Callan}, \bibinfo{person}{Zhiyuan Liu}, {and} \bibinfo{person}{Russell Power}.} \bibinfo{year}{2017}\natexlab{b}.
\newblock \showarticletitle{End-to-End Neural Ad-Hoc Ranking with Kernel Pooling}. In \bibinfo{booktitle}{\emph{Proceedings of the 40th International ACM SIGIR Conference on Research and Development in Information Retrieval}} (Shinjuku, Tokyo, Japan) \emph{(\bibinfo{series}{SIGIR '17})}. \bibinfo{publisher}{Association for Computing Machinery}, \bibinfo{address}{New York, NY, USA}, \bibinfo{pages}{55–64}.
\newblock
\showISBNx{9781450350228}
\urldef\tempurl%
\url{https://doi.org/10.1145/3077136.3080809}
\showDOI{\tempurl}


\bibitem[Xiong et~al\mbox{.}(2017c)]%
        {xiong2017explicit}
\bibfield{author}{\bibinfo{person}{Chenyan Xiong}, \bibinfo{person}{Russell Power}, {and} \bibinfo{person}{Jamie Callan}.} \bibinfo{year}{2017}\natexlab{c}.
\newblock \showarticletitle{Explicit Semantic Ranking for Academic Search via Knowledge Graph Embedding}. In \bibinfo{booktitle}{\emph{Proceedings of the 26th International Conference on World Wide Web}} (Perth, Australia) \emph{(\bibinfo{series}{WWW ’17})}. \bibinfo{publisher}{International World Wide Web Conferences Steering Committee}, \bibinfo{address}{Republic and Canton of Geneva, CHE}, \bibinfo{pages}{1271–1279}.
\newblock
\showISBNx{9781450349130}
\urldef\tempurl%
\url{https://doi.org/10.1145/3038912.3052558}
\showDOI{\tempurl}


\bibitem[Xiong et~al\mbox{.}(2020)]%
        {xiong2020ance}
\bibfield{author}{\bibinfo{person}{Lee Xiong}, \bibinfo{person}{Chenyan Xiong}, \bibinfo{person}{Ye Li}, \bibinfo{person}{Kwok{-}Fung Tang}, \bibinfo{person}{Jialin Liu}, \bibinfo{person}{Paul~N. Bennett}, \bibinfo{person}{Junaid Ahmed}, {and} \bibinfo{person}{Arnold Overwijk}.} \bibinfo{year}{2020}\natexlab{}.
\newblock \showarticletitle{Approximate Nearest Neighbor Negative Contrastive Learning for Dense Text Retrieval}.
\newblock \bibinfo{journal}{\emph{CoRR}}  \bibinfo{volume}{abs/2007.00808} (\bibinfo{year}{2020}).
\newblock
\showeprint[arXiv]{2007.00808}
\urldef\tempurl%
\url{https://arxiv.org/abs/2007.00808}
\showURL{%
\tempurl}


\bibitem[Yamada et~al\mbox{.}(2020)]%
        {yamada-etal-2020-wikipedia2vec}
\bibfield{author}{\bibinfo{person}{Ikuya Yamada}, \bibinfo{person}{Akari Asai}, \bibinfo{person}{Jin Sakuma}, \bibinfo{person}{Hiroyuki Shindo}, \bibinfo{person}{Hideaki Takeda}, \bibinfo{person}{Yoshiyasu Takefuji}, {and} \bibinfo{person}{Yuji Matsumoto}.} \bibinfo{year}{2020}\natexlab{}.
\newblock \showarticletitle{{Wikipedia2Vec: An Efficient Toolkit for Learning and Visualizing the Embeddings of Words and Entities from Wikipedia}}. In \bibinfo{booktitle}{\emph{Proceedings of the 2020 Conference on Empirical Methods in Natural Language Processing: System Demonstrations}}. \bibinfo{publisher}{Association for Computational Linguistics}, \bibinfo{address}{Online}, \bibinfo{pages}{23--30}.
\newblock
\urldef\tempurl%
\url{https://doi.org/10.18653/v1/2020.emnlp-demos.4}
\showDOI{\tempurl}


\bibitem[Yu et~al\mbox{.}(2021)]%
        {hongchien2021ance-prf}
\bibfield{author}{\bibinfo{person}{HongChien Yu}, \bibinfo{person}{Chenyan Xiong}, {and} \bibinfo{person}{Jamie Callan}.} \bibinfo{year}{2021}\natexlab{}.
\newblock \showarticletitle{Improving Query Representations for Dense Retrieval with Pseudo Relevance Feedback}. In \bibinfo{booktitle}{\emph{Proceedings of the 30th ACM International Conference on Information \& Knowledge Management}} (Virtual Event, Queensland, Australia) \emph{(\bibinfo{series}{CIKM '21})}. \bibinfo{publisher}{Association for Computing Machinery}, \bibinfo{address}{New York, NY, USA}, \bibinfo{pages}{3592–3596}.
\newblock
\showISBNx{9781450384469}
\urldef\tempurl%
\url{https://doi.org/10.1145/3459637.3482124}
\showDOI{\tempurl}


\bibitem[Zhang et~al\mbox{.}(2019)]%
        {zhang-etal-2019-ernie}
\bibfield{author}{\bibinfo{person}{Zhengyan Zhang}, \bibinfo{person}{Xu Han}, \bibinfo{person}{Zhiyuan Liu}, \bibinfo{person}{Xin Jiang}, \bibinfo{person}{Maosong Sun}, {and} \bibinfo{person}{Qun Liu}.} \bibinfo{year}{2019}\natexlab{}.
\newblock \showarticletitle{{ERNIE: Enhanced Language Representation with Informative Entities}}. In \bibinfo{booktitle}{\emph{Proceedings of the 57th Annual Meeting of the Association for Computational Linguistics}}. \bibinfo{publisher}{Association for Computational Linguistics}, \bibinfo{address}{Florence, Italy}, \bibinfo{pages}{1441--1451}.
\newblock
\urldef\tempurl%
\url{https://doi.org/10.18653/v1/P19-1139}
\showDOI{\tempurl}


\bibitem[Zheng et~al\mbox{.}(2020)]%
        {zheng-etal-2020-bert}
\bibfield{author}{\bibinfo{person}{Zhi Zheng}, \bibinfo{person}{Kai Hui}, \bibinfo{person}{Ben He}, \bibinfo{person}{Xianpei Han}, \bibinfo{person}{Le Sun}, {and} \bibinfo{person}{Andrew Yates}.} \bibinfo{year}{2020}\natexlab{}.
\newblock \showarticletitle{{BERT-QE}: {C}ontextualized {Q}uery {E}xpansion for {D}ocument {R}e-ranking}. In \bibinfo{booktitle}{\emph{Findings of the Association for Computational Linguistics: EMNLP 2020}}. \bibinfo{publisher}{Association for Computational Linguistics}, \bibinfo{address}{Online}, \bibinfo{pages}{4718--4728}.
\newblock
\urldef\tempurl%
\url{https://doi.org/10.18653/v1/2020.findings-emnlp.424}
\showDOI{\tempurl}


\bibitem[Zhuang et~al\mbox{.}(2023)]%
        {zhuang2023rankt5}
\bibfield{author}{\bibinfo{person}{Honglei Zhuang}, \bibinfo{person}{Zhen Qin}, \bibinfo{person}{Rolf Jagerman}, \bibinfo{person}{Kai Hui}, \bibinfo{person}{Ji Ma}, \bibinfo{person}{Jing Lu}, \bibinfo{person}{Jianmo Ni}, \bibinfo{person}{Xuanhui Wang}, {and} \bibinfo{person}{Michael Bendersky}.} \bibinfo{year}{2023}\natexlab{}.
\newblock \showarticletitle{RankT5: Fine-Tuning T5 for Text Ranking with Ranking Losses}. In \bibinfo{booktitle}{\emph{Proceedings of the 46th International ACM SIGIR Conference on Research and Development in Information Retrieval}} (Taipei, Taiwan) \emph{(\bibinfo{series}{SIGIR '23})}. \bibinfo{publisher}{Association for Computing Machinery}, \bibinfo{address}{New York, NY, USA}, \bibinfo{pages}{2308–2313}.
\newblock
\showISBNx{9781450394086}
\urldef\tempurl%
\url{https://doi.org/10.1145/3539618.3592047}
\showDOI{\tempurl}


\bibitem[Zhuang et~al\mbox{.}(2025)]%
        {zhuang2025rankr1}
\bibfield{author}{\bibinfo{person}{Shengyao Zhuang}, \bibinfo{person}{Xueguang Ma}, \bibinfo{person}{Bevan Koopman}, \bibinfo{person}{Jimmy Lin}, {and} \bibinfo{person}{Guido Zuccon}.} \bibinfo{year}{2025}\natexlab{}.
\newblock \bibinfo{title}{Rank-R1: Enhancing Reasoning in LLM-based Document Rerankers via Reinforcement Learning}.
\newblock
\newblock
\showeprint[arxiv]{2503.06034}~[cs.IR]
\urldef\tempurl%
\url{https://arxiv.org/abs/2503.06034}
\showURL{%
\tempurl}


\end{thebibliography}
\end{document}